\documentclass[iop]{emulateapj}

\usepackage{graphicx}

\newcommand{\ha}{H$\alpha~$}
\newcommand{\hb}{H$\beta~$}
\newcommand{\oiii}{[O {\footnotesize III}]($\lambda\lambda$5007)}

\slugcomment{Accepted for publication in PASP}
\shorttitle{Quantitative Method for the Optimal Continuum Subtraction }
\shortauthors{Hong et al.}

\begin{document}

\title{Quantitative Method for the Optimal Subtraction of Continuum Emission from Narrow-band Images: 
Skewness Transition Analysis}

\author{
Sungryong Hong\altaffilmark{1,2},
Daniela Calzetti\altaffilmark{1},
and
Mark Dickinson\altaffilmark{2}
}

\altaffiltext{1}{Department of Astronomy, University of Massachusetts,  Amherst, MA 01003}
\altaffiltext{2}{NOAO, 950 N. Cherry Avenue, Tucson, AZ 85719, USA}


\begin{abstract}
We present an objective method to remove the stellar continuum
emission from narrow--band images to derive emission--line
images. The method is based on the skewness of the pixel histogram of the residual images. 
Specifically, we exploit a transition
in the skewness of the signal in the continuum--subtracted image,
which appears when the image changes from being under--subtracted to
over--subtracted. 
Tests on one--dimensional artificial images
demonstrate that the transition identifies the optimal scaling factor
$\mu$ to be used on the broad-band image I$_B$ in order to produce the
optimal line--emission image I$_E$, i.e., I$_E$=I$_N$-$\mu$ I$_B$,
with I$_N$ the original (un--subtracted) narrow--band image. The
advantage of this method is that it uses all information--bearing
pixels in the final image, and not just a sub--set of those pixels
(the latter being common in many traditional approaches to stellar continuum
removal from narrow--band images). We apply our method to actual images,
both from ground--based and space facilities, in particular to WFPC2 and
ACS images from the Hubble Space Telescope, and we show that it is
successful irrespective of the nature of the sources (point-like or extended).
We also discuss the impact on the accuracy of the method of non--optimal
images, such as those containing saturated sources or non--uniform
background, and present `workarounds' for those problems.

\end{abstract}
\clearpage

\clearpage
\clearpage

\section{Introduction}

Investigations of emission lines from astronomical sources provide
a host of information from those sources, including the chemical,
kinematic, and physical conditions. While spectroscopy is generally
preferred when analyzing a single or a few sources, narrow--band imaging
has traditionally offered the advantage of surveying extended regions
of the sky in one or a few emission lines of interest.

One of the most difficult steps when deriving the intensity or equivalent
width (EW) of an emission line is the determination
of the underlying continuum level to be subtracted off the total
intensity at the wavelength of the emission to produce a line--only signal.
For narrow--band images, this step is usually accomplished by obtaining
a second image, either in a broad--band filter (thus, enhancing the signal
of the underlying continuum relative to that of the emission line) or
in an emission--line--free narrow--band filter, at a wavelength adjacent
to that of the emission line, and subtracting a scaled version of this
image from the original narrow--band image. The scaling parameter $\mu$
is most often determined from (1) measurements of emission--line--free sources
in the same image (e.g., Helou et al. 2004, Calzetti et al. 2007), 
or from (2) ratios of the transmission efficiency of the filters used in the two images, 
or a combination of the two methods (Kennicutt et al. 2008).

Either approach (1) or (2) can give unsatisfactory results: the use of
emission--line--free sources exploits only a subset of the available signal
in the image, while the use of the filters transmission curve ratio requires
accurate a--priori knowledge of both the filters characteristics and the
spectral energy distribution of the sources of interest. Additionally,
the first approach introduces an element of subjectivity in the determination
of which sources are free of emission lines.

We discuss in this paper a method for subtracting the underlying continuum
from narrow--band images that attempts to remove much of the subjectivity
or uncertainty built into other methods. We show, in what follows, that
the skewness of the signal distribution in the pixels of the
continuum--subtracted image is a sensitive indicator of the optimal value
for the scaling factor $\mu$, under a large range of characteristics for
the input images.
In \S2 we present a simple simulation model to show 
how the optimal scaling factor is related to a specific feature (that we call ``transition'') 
in the skewness trend in the residual (continuum--subtracted) images.
Then, we show two typical applications of this method to the images from the Advanced Camera for Survey(ACS) of 
the Hubble Space Telescope (HST), and the optical data of the Spitzer Infrared Nearby Galaxies 
Survey (SINGS, Kennicutt et al. 2003) to include examples of both space-based and ground-based data. 
In \S 3 we present several examples of anomalous transitions and show how to fix such anomalies. 
From the implications of our simulations and those anomalies, we derive a set of criteria for the range of applicability 
of our method.  
In \S 4 we summarize the results and discuss pathways for developing user--friendly routines and/or interfaces that enable using 
this method efficiently on astronomical images.

\section{Method Description: skewness transition and the optimal subtraction}
Our skewness transition method is based on a simple observation: 
{\em near the optimal subtraction, skewness values of residual images 
show a transitional behavior and the center of the transition is the optimal subtraction.} 
For the continuum subtraction problem, a common experience is that 
over-subtracted images will tend to have a large number of negative-valued pixels, 
while under--subtracted images will show excess flux resulting in more positive-valued pixels 
than the optimally subtracted image. 
This general property can be expressed as an ``asymmetry'' or skewness of pixel histogram in the image, 
which we can exploit for our optimization problem. 
The skewness of a distribution is defined as:  
\begin{equation}
skewness = \frac{1}{N-1} \sum_{i=1}^{N} \big( \frac{x_i - m}{\sigma}  \big)^3 
\end{equation}
where $m$ is a mean, $\sigma$ is a standard deviation, and $N$ is the sample number.  
The skewness is a direct indicator of asymmetry. Symmetric functions such as a Gaussian have skewness = 0. 
If a function has a long positive tail (many pixels with excess flux), 
it has a positive skewness and called positive-skewed; a negative-tail in the pixel value distribution indicates a negative skewness. 

Pixel histograms of ``sky'' images in astronomy generally possess Gaussian or Poisson distributions, which are, 
therefore, symmetric or slightly positive-skewed. This means that positive skewness values of real images are 
mostly due to astronomical signals. If we assume that 
the signal is composed of the background (sky), the stellar continuum, and the line-emission,  
then our problem is to remove both the sky and the stellar continuum from the mixed observed signal 
while preserving the emission line portion of it. 
When we subtract some continuum from the original image, the overall signal strength decreases. 
This means the skewness decreases as we subtract more continuum from the original image. 
As we continue subtracting continuum from the original image and move toward an over-subtracted image, 
the skewness transitions from positive stellar residual to negative stellar residual. 
If the stellar component dominates over the line emission component, the transition occurs near the skewness = 0; 
i.e. the skewness transits from positive to negative.  
The question becomes  whether we can exploit the transition point between under- 
and over-subtraction to obtain an optimal continuum-subtraction algorithm. 

We investigate the meaning of such empirical transition in a 
controlled ``experiment'' by building a simple model and 
present the implications from it in the following section. 
We then move to actual applications to real astronomical images in the next section.

\subsection{Implications from simulation}
The purpose of this section is to show the results of a simple simulation model and 
to draw some important implications about the skewness transition and its relation 
with the optimal continuum subtraction. The model also provides criteria 
for application of the method to real images and some clues to resolve anomalous 
cases.

\subsubsection{One dimensional model}

While images are two dimensional scalar data sets of x, y, and pixel value, 
the statistics on them usually compresses the spatial dimensions and 
only deal with the pixel values. Hence, the dimensionality of our image 
is not important to study its statistical properties. 
In addition, most tasks reading pixel values from images follow row-by-row 
or column-by-column directions virtually treating images as one dimensional data. 
It is, therefore, reasonable to treat images as one dimensional arrays for our skewness method.  


\subsubsection{Definitions}

We assume that a narrow-band image $I_N(x)$ consists of four components, extended 
line-emission $E(x)$, emission from stars $\sum S_{i}(x)$, background $B_{N}$, and combined noise 
of all kinds $\sigma_{N}(x)$ including Poisson (or Gaussian) noise of each signal, dark current, and readout noise. 
We also assume that a broad-band image $I_B(x)$ consists of three components:  
stellar emission $\sum S'_{i}(x)$, background $B_{B}$, and combined noise $\sigma_{B}(x)$, and 
any other components are negligible for each image. Those are written mathematically as: 
\begin{eqnarray}\label{eqn:ab}
I_N(x) & = & E(x) + \sum S_{i}(x) + B_{N} + Noise[\sigma_{N} (x)] \\
I_B(x) & = & \sum S'_{i}(x) + B_{B} + Noise[\sigma_{B} (x)]  \\
\sigma_{N} & = & \sqrt{\sigma_{E}^2 +\sigma_{S}^2 + \sigma_{B_N}^2 + dc_{N} + RN_{N}^{2}} \\
\sigma_{B} & = & \sqrt{\sigma_{S'}^2 + \sigma_{B_B}^2 + dc_{B} + RN_{B}^{2}} 
\end{eqnarray}
where $x$ is the pixel coordinate, $\sigma_E^2$ is the variance for $E(x)$, 
$\sigma_{S}^2$ and $\sigma_{S'}^2$ for stellar emissions, 
$\sigma_{B_N}^2$ and $\sigma_{B_B}^2$ for backgrounds, $RN_{N}^2$ and $RN_{B}^2$ for read-out noises, and 
$dc_N$ and $dc_B$ are dark currents. 

To quantify the amount of subtraction, we define a residual image, $R(x; \mu)$, as  
\begin{eqnarray}
R(x; \mu) & \equiv & I_N(x) - \mu I_B(x) \nonumber \\
& = & E(x) + (B_{N} - \mu B_{B}) + \sum (S_{i}(x) - \mu S'_{i}(x))  \nonumber \\
& &  +~ Noise[\sqrt{\sigma_{N}^2 (x) + \mu^2 \sigma_{B}^2 (x)}] \nonumber \\
 & = & E(x) + \Delta_{B}(\mu) + \Delta_{S}(x; \mu) + Noise[\sigma_{SUB}(x;\mu)]
\end{eqnarray}
\begin{eqnarray}
\Delta_{B}(\mu) & \equiv & B_{N} - \mu B_{B} \\
\Delta_{S}(x; \mu) & \equiv & \sum (S_{i}(x) - \mu S'_{i}(x)) \\
\sigma_{SUB}(x; \mu) & \equiv & \sqrt{\sigma_{N}^2 (x) + \mu^2 \sigma_{B}^2 (x)} 
\end{eqnarray}
where $\mu$ is a parameter to control the amount of subtraction, $\Delta_{B}(\mu)$ is a background residual function, 
$\Delta_{S}(x; \mu)$ is a stellar residual function, and $\sigma_{SUB}^2(x; \mu)$ is the variance of all the combined noises. 
The residual functions are the terms to describe the intrinsic differences of  
backgrounds and stellar continuums in the two images. 

Finally, we define the skewness function, $s(\mu)$, by measuring the skewness 
of the residual image for a given $\mu$.
\begin{equation}
s(\mu) \equiv skewness(R(x; \mu)) \label{eq:skew}, 
\end{equation}
where the definition of skewness is presented in Equation 1. 
In this mathematical framework, the continuum subtraction problem is rephrased as a problem to 
find an optimal $\mu$ parameter which minimizes the stellar residual function.

\subsubsection{Numerical realization}

In this section, we assign simple numerical functions to each term in the previous equations. 
We simulate images with arrays of $10^5$ pixels, which is roughly a $300\times 300$ 2D image. 
The size is smaller than a typical image but large enough 
to mimic most statistical behaviors. We define a smooth and extended function for E(x) and 
take a Gaussian profile to represent each star as written below,  
\begin{eqnarray}
E(x) & = & 7.0 \times e^{-\frac{(x-50000)^{2}}{2\times 10000^{2}}} \nonumber \\
	& & \times \sin^2(x/500) \times \sin^2(x/700)\label{eq:source} \\
\sum_{i=1}^{N} S_i(x) & = &\sum_{i=1}^{N} a_i e^{-\frac{(x-b_i)^{2}}{2 c_i^{2}}}  \\
\sum_{i=1}^{N} S'_i(x) & = &\sum_{i=1}^{N} d_i e^{-\frac{(x-e_i)^{2}}{2 f_i^{2}}}  
\end{eqnarray}
where $N$ is a number of stars. We choose a basic set of parameters, 
\begin{eqnarray}
N & = & 200 \nonumber \\
a_{i} & = & randomu[0,200]+2.0 \nonumber \\
b_{i} & = & randomu[0,10^5] \nonumber \\
c_{i} & = & 5.0 \nonumber \\
d_{i} & = & 4\times a_{i} \nonumber \\
e_{i} & = & b_{i} \nonumber \\
f_{i} & = & c_{i} \nonumber \\
Noise[\sigma_{N} (x)] & = & randomn(\sigma = 1)  \nonumber \\
Noise[\sigma_{B} (x)] & = & randomn(\sigma = 2) \nonumber \\
B_{N} & = & -10 \nonumber \\
B_{B} & = & -10 \nonumber 
\end{eqnarray}
where ``$randomu[a,b]$'' chooses a random number uniform between `a' and `b',  
``$randomn[\sigma=c]$'' produce a Gaussian distribution with the sigma value of `c'. We refer  
this set of parameters as our ``Reference Set".

$E(x)$ is chosen to be smooth and extended throughout the whole array, 
which we model as a combination of a exponential and a trigonometric function. 
Many other functional forms could be used. 
$a_i$ and $d_i$ represent the fluxes of simulated stars. The fluxes are randomly chosen from 2 to 202. 
$b_i$ and $e_i$ represent the positions of simulated stars. They are chosen randomly throughout the whole array. 
$c_i$ and $f_i$ represent the width of Gaussian PSFs and we fix the value at 5.0 for both images. 
The noise components are chosen as Gaussian distributions with the given sigma value. 
In the Reference Set, the two stellar components, $S_{i}(x)$ and $S'_{i}(x)$, 
are equal except the flux scale, $d_{i}=4\times a_{i}$. 
All other parameters are set equal in each image, without registration errors, PSF mismatches, 
or color and background variations.  
Therefore, the stellar residual function vanishes at $\mu=0.25$; 
$\Delta_{S}(x; \mu=0.25)=0.0$ in the Reference Set, which corresponds to the perfect continuum subtraction.

\subsubsection{Symmetric transition vs. Asymmetric transition for the skewness}

\begin{figure*}[t]
\begin{center}
\includegraphics[width=0.99\textwidth]{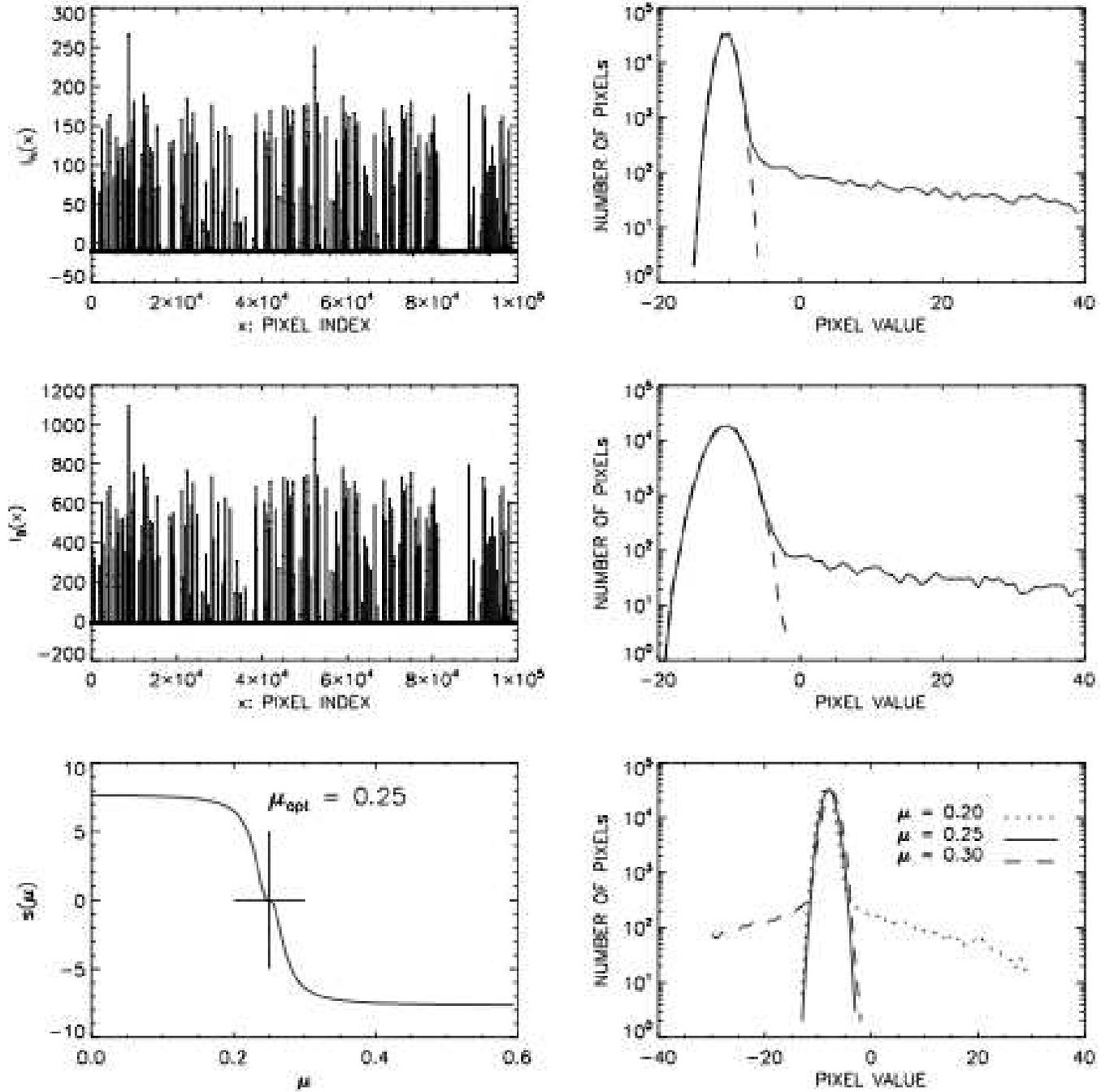}
\end{center}
\caption{Basic results for the Reference Model. We can observe that the skewness function is symmetric 
with a transition feature at $\mu=0.25$. This is the key feature of our continuum subtraction method. 
}\label{fig:refer1}
\end{figure*}

Here we show the skewness values of residual images for our numerical models. 
Before we investigate more practical cases with a model, we present a trivial case called the ``Reference Model". 
In the Reference Model, we set $E(x) = 0$ and take the Reference Set values for all other parameters. 
Hence, the two simulated images, $I_N(x)$ and $I_B(x)$, are identical at $\mu=0.25$. 
Figure~\ref{fig:refer1} shows the results of the Reference Model. 
The top-left (middle-left) panel shows the simulated narrow-band image (broad-band image). 
The corresponding right panel shows the pixel histogram for each simulated image. The bottom-left figure 
shows the skewness function for the Reference Model and the bottom-right panel shows the pixel 
histograms for three different cases:  
under-subtracted ($\mu = 0.20$), optimally-subtracted ($\mu = 0.25$), and over-subtracted ($\mu = 0.30$) images. 
Basically the skewness values are monotonically decreasing as we increase the subtraction scaling factor between 
the two images. This monotonically decreasing trend exhibits a ``flattening'' effect near the optimal subtraction, 
because the skewness value contributed by stellar component is zero at the optimal subtraction. 
For the Reference Model, the residual histogram is purely Gaussian at $\mu = 0.25$ due to 
the absence of a line-emission term ($E(x)=0$) ; hence the skewness is zero at the perfect (optimal) subtraction. 
We can also observe that the skewness function $s(\mu)$ is symmetric
\footnote[1]{In mathematical terms, it is `antisymmetric'. We call both of antisymmetry and symmetry as `symmetric' for simplicity, 
because we are only interested in the difference between symmetry and asymmetry.}
at ($\mu=0.25, s(\mu = 0.25) = 0.0$) 
with a \emph{transition feature} by which we can identify where the axis of symmetry is located. 
The transition and its symmetric shape are the key feature of our skewness transition method 
for finding the optimal continuum subtraction. For a smooth profile of $s(\mu)$, 
the transition point is located at the inflection point of $s''(\mu) = 0$. 

\begin{figure*}[t]
\centering
\includegraphics[height=7.0 in]{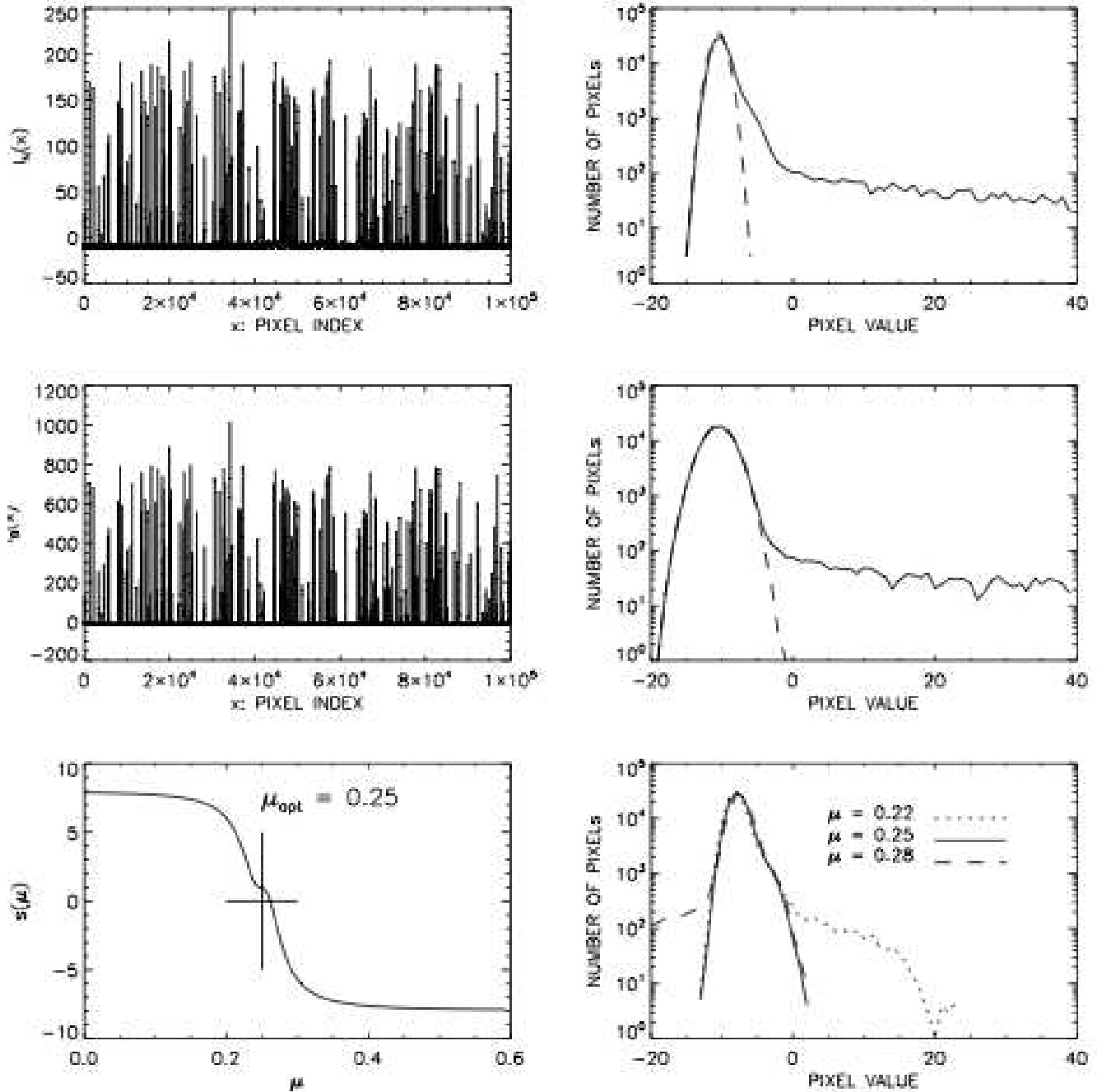}
\caption{The same as Figure~\ref{fig:refer1}, but for the Practical Model. The transition is still (locally) symmetric 
at the exact solution, $\mu=0.25$ (bottom-left panel). By locating the center of the transition, we can find the optimal choice 
of subtraction factor $\mu$.
}\label{fig:practical2}
\end{figure*}

Now we add the emission-line component $E(x)$ to the narrow-band image, $I_N(x)$. The $E(x)$ term of Equation~\ref{eq:source} is 
defined to be faint enough, in flux, 
to minimally change the symmetric transition. Figure~\ref{fig:practical2} shows the results of this model. 
We call this our ``Practical Model'', because this will be the most general case in our practical applications. 
The notable differences between the Practical Model and the Reference Model are :
\begin{enumerate}
\item The skewness is non-zero at the optimal subtraction value, $s(\mu = 0.25) \neq 0$, because of non-zero line-emission.  
\item The transition is slightly asymmetric at $\mu = 0.25$, but still we can locate the center of transition corresponding to optimal subtraction.
\end{enumerate}

\begin{figure*}[t]
\centering
\includegraphics[height=3.2 in]{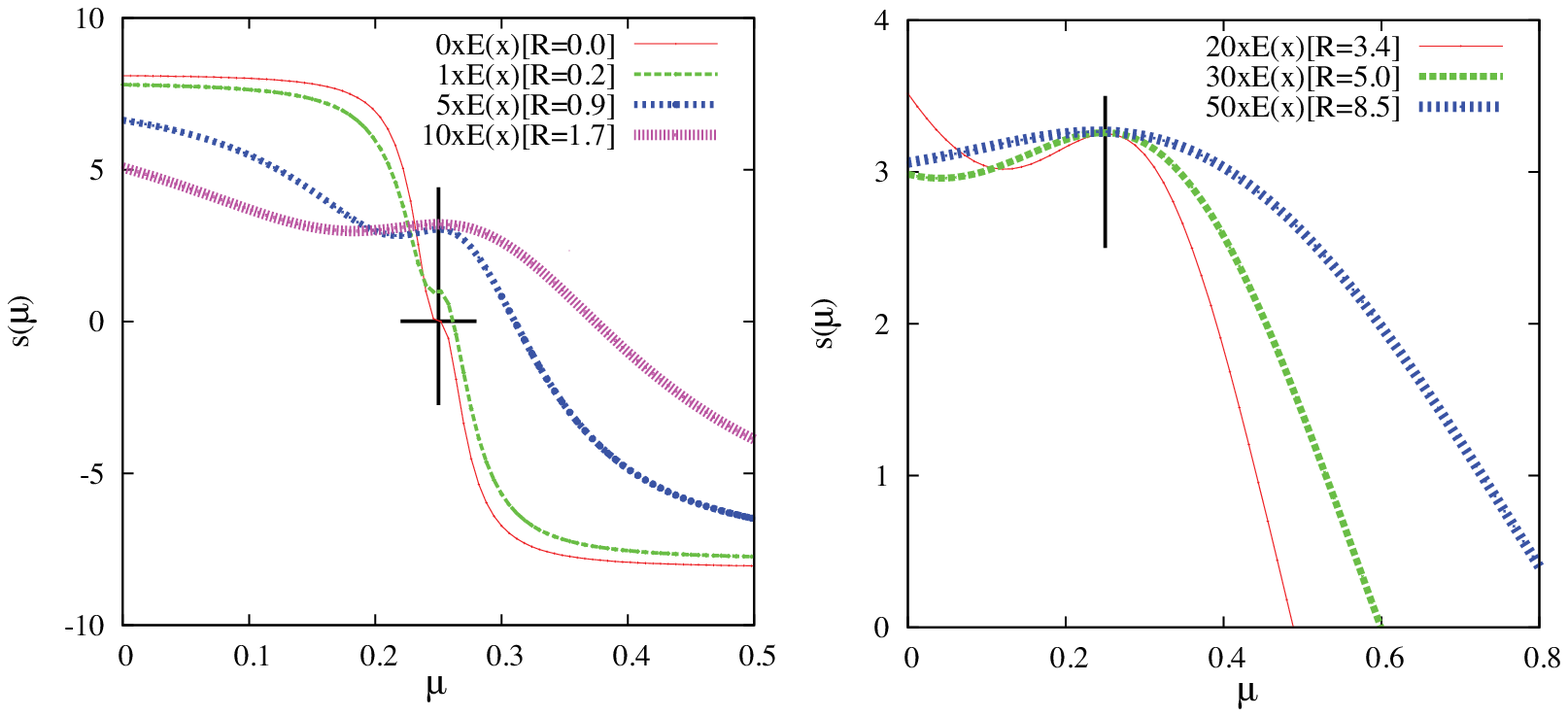}
\caption{The skewness functions, $s(\mu)$, for various line-emissions. The ratio, R, is a ratio of total fluxes 
defined as $R \equiv \int E(x) dx / \int \sum S_i(x) dx$. 
For $R < 0.5$, the transition is relatively symmetric. We can determine the center of transition with high accuracy. 
For $0.5 < R < 1.0$, the transition is asymmetric but still we can locate the transition point. 
For $R > 1.0$, most of skewness is contributed by the line emission term $E(x)$. So the skewness variation before 
the transition is minor. Their trends of $s(\mu)$ are flat or slightly increasing near the transition. After the transition, the skewness 
decreases monotonically. Hence, we still make a rough guess using the decreasing pattern for the $R >> 1.0$ regime. 
Figure \ref{fig:ngc4449fourth13} shows the effect of this more extreme ``R" dependence. 
}\label{fig:ex3}
\end{figure*}

The effect of adding the line-emission $E(x)$ component is to increase the asymmetry of the transition feature. 
Figure~\ref{fig:ex3} shows the trend of the asymmetric behavior of $s(\mu)$ 
by increasing the line-emission in the narrow-band image. 
The ratio, R, is a ratio of total fluxes 
defined as $R \equiv \int E(x) dx / \int \sum S_i(x) dx$. This represents a relative strength between stellar component and line emission component. 
For $R < 0.5$, the transition is relatively symmetric. We can determine the center of transition with high accuracy. 
For $0.5 < R < 1.0$, the transition is asymmetric but still we can locate the transition point. 
For $R > 1.0$, most of skewness is contributed by the line emission term $E(x)$. So the skewness variation before 
the transition is minor. Their trends are flat or slightly increasing near the transition. After the transition, the skewness 
decreases monotonically. Due to the dominance of line emission, the continuum subtraction is less important for this case. 
Also, we can still make a rough guess using the decreasing pattern for the $R >> 1.0$ regime shown in Figure~\ref{fig:ex3} and \ref{fig:ngc4449first10}.

This experiment provides the first general recipe for our approach: the optimal subtraction 
value will be more accurately recovered if we can choose a section of our image 
where the stellar emission dominates over the line emission across the selected image 
section. 

\subsubsection{Error tolerance}

The previous models are ideal cases because we set most of the uncertainties to zero. 
In this section, we present the effect of adding different backgrounds, noise levels, 
registration errors, PSF mismatches, and stellar flux errors  
on the transition feature of the Practical Model. 

\begin{figure}[t]
\begin{center}
\includegraphics[width=1.0\columnwidth]{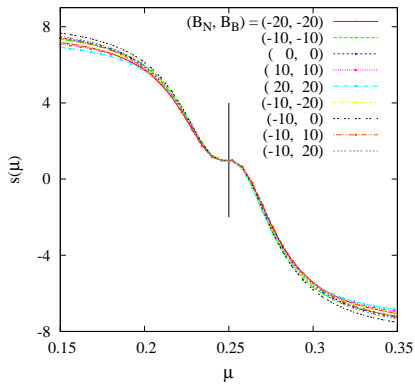}
\end{center}
\caption{The skewness functions for the Practical Model for various background combinations for narrow-band and broad-band images. 
}\label{fig:back4}
\end{figure}

\begin{figure}[t]
\centering
\includegraphics[width=1.0\columnwidth]{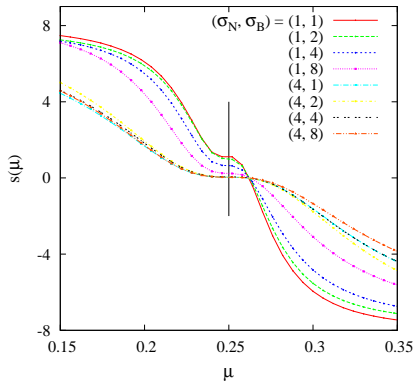}
\caption{The skewness functions for various sigma values of Gaussian noises for narrow-band and broad-band images. 
}\label{fig:sig5}
\end{figure}

For the background, we can easily predict that differences in backgrounds between narrow-band and broad-band images 
do not affect the skewness value, since the background of an image is 
a constant generally related to the first moment of the distribution. 
Figure~\ref{fig:back4} shows the skewness functions of many different background combinations 
for narrow-band and broad-band images. The small differences for various background combinations 
are caused by the removing process of deviant pixels, ``iterstat'', described in the following section. 
From the figure, we infer that we can generally ignore the effect of different backgrounds 
between narrow-band and broad-band images on the transition feature of $s(\mu)$. 
For the noise levels, Figure~\ref{fig:sig5} shows the skewness functions for various $\sigma_{N}$ and $\sigma_{B}$ combinations. 
Asymmetries can develop for some combinations, but overall the optimal solution is at the center of the transition.

\begin{figure*}[t]
\begin{center}
\includegraphics[height=5.5 in]{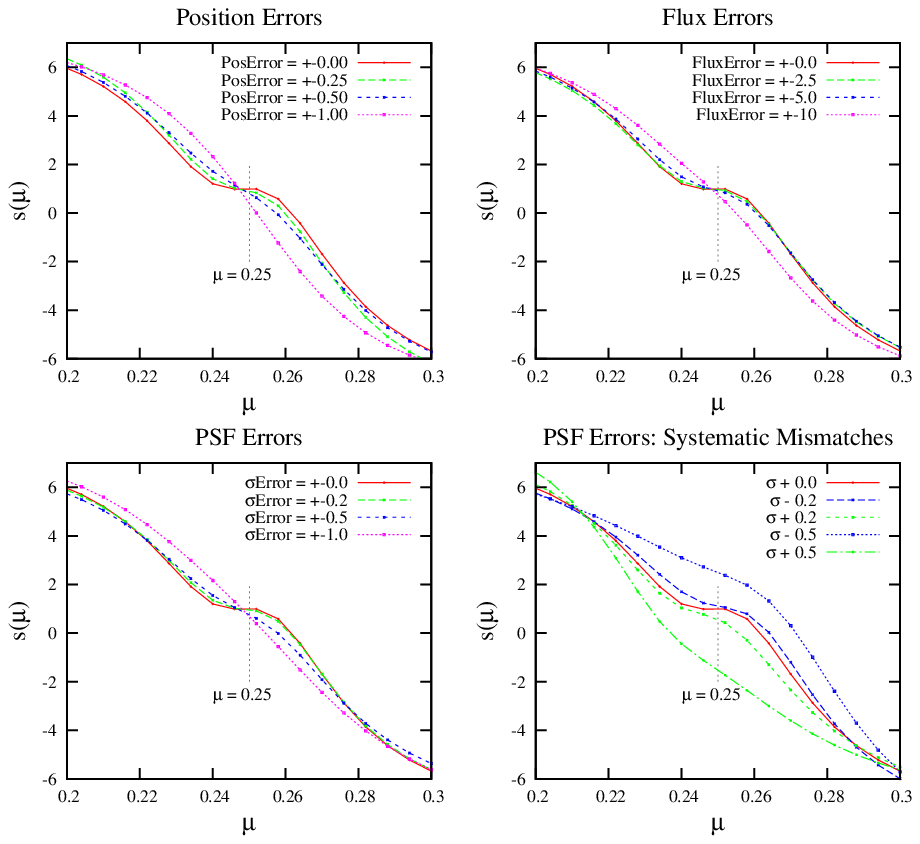}
\end{center}
\caption{The skewness transitions resulting for various errors added to the images. 
The PSF has $\sigma = 5$ in the Practical Model used for this figure. So PosError = 0.5, $\sigma$Error = 0.5, and 
$\sigma$ offset = 0.5 are 10 \% of the PSF sigma value. Since the peaks of stellar flux are uniformly distributed between 
2.0 to 202.0 in this model, FluxError = 10 is 10 \% of the median of the stellar fluxes. 
}\label{fig:errors6}
\end{figure*}

Figure~\ref{fig:errors6} shows the effects of adding position errors (top-left panel), 
flux errors (top-right panel), 
and PSF errors (bottom panels) on the Practical Model. 
The errors illustrate the types of effects that can arise from 
inaccurate registration (position errors) of the narrow and broad band images; 
the flux errors from flat-fielding inaccuracies and different colors of the two bandpasses; 
and the PSF errors from different seeings and inaccuracies of PSF matching. 

For the position errors (or registration error), we add uniform random numbers 
to the positions of stellar objects with the given range of ``PosError'' in the simulated broad-band image, 
thus simulating shifts in pixels.  
The position errors change the shape of the transition to more negative-sloped ones 
and eventually smear out the transition for a $\pm 1.0$ shift, which 
corresponds to 20\% of the sigma width, 5, of the Gaussian PSF in the Practical Model. 
In a similar sense, we add uniform random numbers to the fluxes and the widths 
of the stellar PSFs to the broad-band images. 
Both show the same result that the transition is conserved while a certain amount of error is added  
and, after that threshold, the transition is smeared out. The FluxError $\pm 10$ and 
the PSF width error $\sigma Error = \pm1.0$ correspond to 10\% of average stellar flux and 20\% of 
the width of Gaussian PSF. The difference in color and dust extinction between the two filters 
can produce a flux difference between the two images similar to the one we simulate. 
The widths of PSFs also show a certain amount of scatter in actual images. 
These practical issues can be considered as flux errors and PSF errors. 

A systematic difference of the size of PSFs can occur when using different instruments for the two images. 
It is possible that there can be a systematic offset even after matching the PSFs. 
To investigate the effect of a systematic difference in the PSF size, 
we add a constant to the width of the Gaussian PSF in the simulated broad-band image. 
In this case, since the profile shapes are different for the two images, 
the stellar residual function does not vanish at $\mu = 0.25$. 
The bottom-right panel of Figure~\ref{fig:errors6} shows the transitions for various systematic mismatches. 
For broader profiles in the broad-band image (positive additions to the $\sigma$ values), the transition 
shifts to a smaller value of $\mu$, because, to minimize the difference between the two profiles, 
the peak of the broader profile has to be lowered. 
For sharper profiles (negative additions to the $\sigma$ values) in the broad-band image, the transition is shifted to larger $\mu$. 
As other errors, the transition is smeared out when the offset is too large. 

To summarize, we have the two important implications from investigating the effects of errors on the transition:   
\begin{enumerate}
\item The transition is conserved for errors in source registration, PSF width, or flux scaling that are smaller than $\sim$ 10 \% $-$ 20 \%. 
\item Most errors make the slope of the transition more negative (we call this ``down-slope transition''). 
\end{enumerate}
The smearing-out limits in our model are just guidelines since our model is simplistic and 
its purpose is to provide theoretical guidance not to simulate accurate features in the continuum subtraction of real images. 
In most practical applications, the error tolerance is robust enough to find the transitions in most of the subtracted images. 
Most transitions in practical applications are down-sloped because most uncertainties are inevitable.

\subsubsection{Removing deviant pixels}

Real images contain many cosmetic defects including unmasked cosmic rays and saturated pixels. 
Such pixels can dominate the skewness value even though the number of the pixels is small. 
Therefore, we need a procedure to remove such deviant pixels. 
In our approach, we adopt {\it iterstat.cl}\footnote[2]{http://stsdas.stsci.edu/cgi-bin/gethelp.cgi?iterstat.src} from the  
Image Reduction and Analysis Facility(IRAF)
\footnote[3]{IRAF is distributed by the National Optical Astronomy Observatory, 
which is operated by the Association of Universities for Research in Astronomy (AURA) 
under cooperative agreement with the National Science Foundation.
}and 
{\it djs\_iterstat.pro}\footnote[4]{http://www.lancesimms.com/programs/IDL/idl\_H4RG/djs\_iterstat.pro. A Python port is available at http://www.lancesimms.com/programs/Python/functions/Djs\_Iterstat.py.} 
from the Interactive Data Language(IDL). 
Both scripts have two parameters, SIGREJ and MAXITER. 
They iteratively recalculate the statistics ignoring the outliers outside of $m_i \pm$SIGREJ$\times \sigma_i$, 
where $m_i$ and $\sigma_i$ are mean and standard deviation at the $i$-th step, 
and stop when one of these conditions is met: 
(1) The maximum number of iterations, as set by MAXITER, is reached. 
(2) No new pixels are rejected, as compared to the previous iteration. 
(3) At least 2 pixels remain from which to compute statistics. 
Though we adopt the iterstat procedure for rejection, 
that is not the only option to remove the deviant pixels. 
The rejection algorithm is needed to prevent deviant pixels from dominating the skewness measurement,  
and any other routine that has this purpose can be used on the images.

\begin{figure}[t]
\centering
\includegraphics[width=1.0\columnwidth]{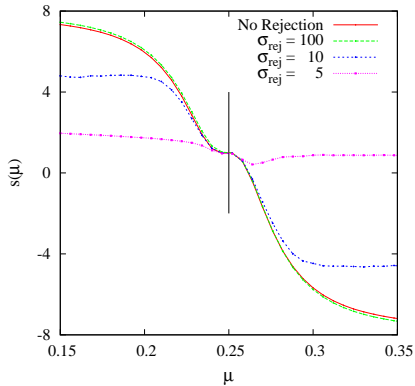}
\caption{The skewness function for various rejection thresholds, $\sigma_{rej}$, using the  {\it iterstat} routine. 
Basically implementing a rejection algorithm on the data preserves the transition. 
}\label{fig:iter7}
\end{figure}

Figure~\ref{fig:iter7} shows the effect of the iterstat routine. We set the maximum number of iterations  
to MAXITER=10. By decreasing the rejection threshold, $\sigma_{rej}$, we remove more pixels from the residual images. 
In our simulation models, the transitional feature does not change until we lose most of the stellar flux. 
So basically the rejection procedure, iterstat, conserves the transition. 
In practical uses, the rejection algorithm can amplify other possible errors, such as registration errors and PSF errors, 
and, eventually, smear out the transition. Therefore, theoretically the rejection process does not change the transition 
but practically strong rejection criteria can affect the transition. 

\subsubsection{Summary: Implications from simulation}

In this section, we have presented the results of our one dimensional simulations. We have shown that there is a transition 
in the skewness values of the  residual images. And the location of the transition provides the optimal value $\mu$ 
for stellar flux removal. The line emission term, $E(x)$, 
increases the asymmetry of the transition. Without the line emission term, we 
have a simple symmetric transition as shown in the Reference Model. To keep the transition in near symmetric shape, 
we need to choose a section of the image\footnote[4]{Generally a section of image needs to be larger than $500\times500$. 
Since we use all the pixels in the section, we do not lose our strength of statistical completeness.}
 where the stellar flux dominates over the emission line flux, $R < 0.5$ (section~2.1.4), for 
high accuracy. The transition is error-resistant for large ranges in the possible observational uncertainties. 
The {\it iterstat} rejection routine we adopt to remove deviant pixels (CRs and saturated pixels) 
conserves the transition, provided the integrity of the image (in terms of total stellar flux) is preserved.

Our skewness transition method is better for statistical sampling than 
any extant methods, such as choosing several point-source objects 
to measure flux ratios, because we use the whole pixel distribution in a section of (or an entire) image. 
For the stellar dominant regime, $R < 0.5$, we can make an accurate estimation for subtraction. 
Computation time is also relatively short, because we only measure a skewness for each residual image. 
The only weak point on this method is that we may not find the transition in some pathological cases. 
In the following sections, we will investigate our method on real images 
and also present how to deal with pathological cases.

\subsection{Applications to real images}

The previous section shows the expectations from models for where and how the 
skewness method works. But our simulated images are too simplistic 
to represent real images. 
In this section we will present typical examples which demonstrate that the expectations from 
models are applicable to real data and 
that the optimal $\mu$ value is found at the center of the transition 
for stellar dominated images.

\subsubsection{Case 1: Ground-based observations}
\begin{figure*}[t]
\centering
\includegraphics[height=4.8 in]{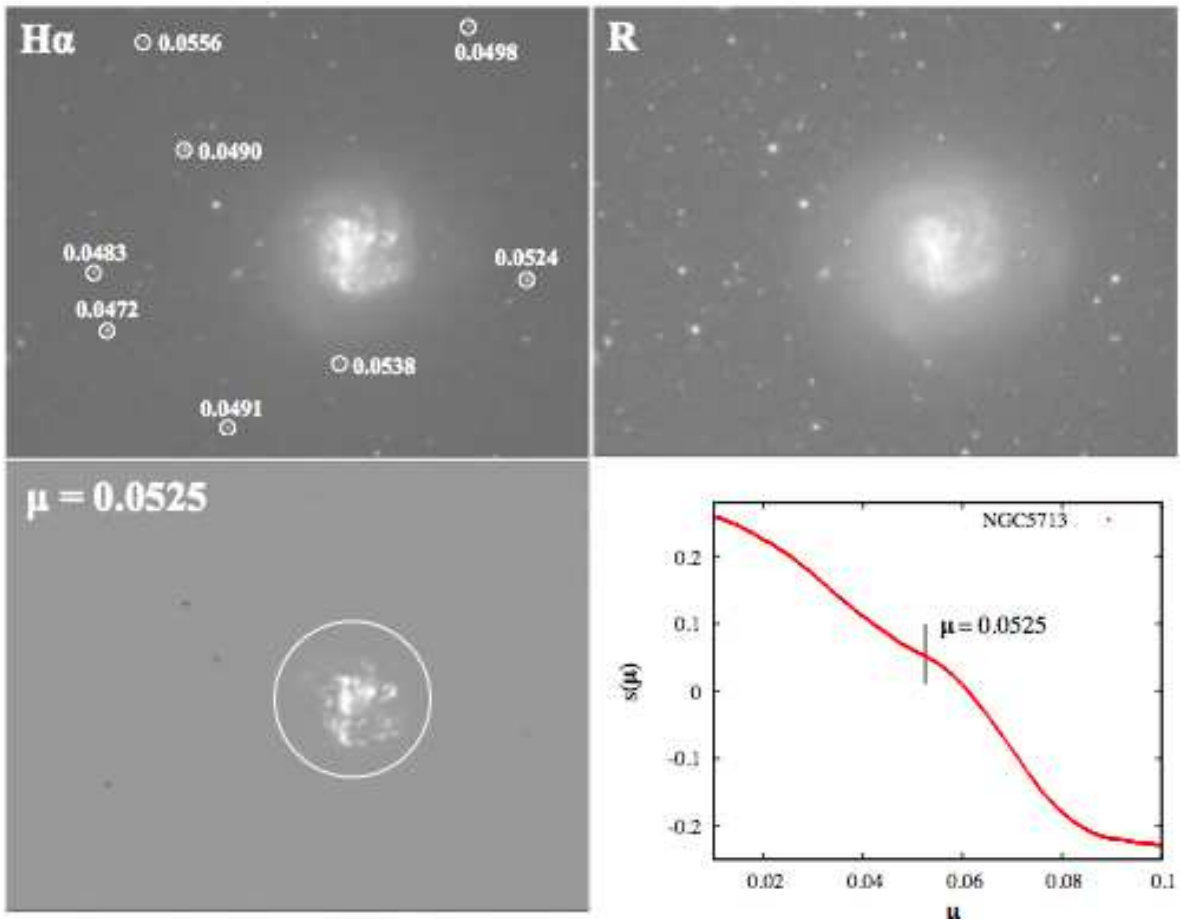}
\caption{The H$\alpha$ image(top-left) and the R-band image(top-right) for NGC5713,  
observed with the 1.5 m telescope at CTIO. 
We select several stars marked with open circles  and list the flux ratios 
between the two images in the top-left image.
By  using this more traditional approach to stellar continuum subtraction, 
we would get a scaling factor $\mu = 0.049$ for the broad-band image.  
The bottom-right panel shows 
the skewness function. The transition is slightly asymmetric but enough to locate 
where the optimal transition is. We choose $\mu = 0.0525$ for the optimal subtraction. 
The bottom-left panel shows the continuum-subtracted image using $\mu=0.0525$; 
perfectly subtracted foreground stars in this image support our choice $\mu = 0.0525$ for optimal subtraction.  
The open circle covering the galaxy shows the aperture size used for comparing aperture photometry.}
\label{fig:ngc5713first8}
\end{figure*}

Figure \ref{fig:ngc5713first8} shows the H$\alpha$ image (top-left) and the R-band image (top-right) for NGC5713 
observed with the 1.5 m telescope of Cerro Tololo Inter-American Observatory (CTIO). 
The data are part of the Spitzer Infrared Nearby Galaxies Survey (SINGS, Kennicutt et al. 2003). 
To obtain a continuum-free \ha image, we use the R-band image 
to remove the stellar component from the narrow-band image. 
One common method is to measure the flux of stellar objects in the two images 
and take the ratios of them for the scaling factor of continuum subtraction. 
The 8 open circles in the \ha image indicate the stars we choose to calculate the flux ratios (the number adjacent to each circle) 
between the narrow-band and the broad-band images. We use the IRAF/IMEXAMINE task for photometry of each star. 
The minimum ratio among our chosen stars is 0.047 and the maximum ratio is 0.056. 
From those ratios, we can point out the two caveats for this continuum subtraction method; 
(1) Stars are located in the foreground of the target. 
(2) The distribution seems to be bimodal clustered around 0.049 and 0.053. This is a selection bias 
caused by choosing several bright stars in the image. 
 
The bottom-right panel in Figure \ref{fig:ngc5713first8} shows the skewness function obtained with the iterstat routine 
($3\sigma$ rejection, 10 iterations; the rejection thresholds of $3\sigma$ or $5\sigma$ are suitable for most 
practical cases. Our simulation has somewhat higher $\sigma$ thresholds because of our simplistic 
implementations for stellar and extended sources). This is a typical example consistent with our theoretical results. 
All of the flux ratios from the stellar objects are scattered around the center of the transition. 
The overall shape of the skewness function is quite similar to that of our Practical Model. 
The slope of the transition is negative as our error-tolerance test suggests. 
The transition is slightly asymmetric but enough to locate the optimal position. 
We choose $\mu = 0.0525$ for the optimal location, 
and we show the residual image for $\mu = 0.0525$ in the bottom-left panel of Figure \ref{fig:ngc5713first8}. 
In this image, the foreground stars are almost perfectly subtracted, supporting our choice for the optimal value of $\mu$. 

\begin{figure*}[t]
\centering
\includegraphics[height=4.8 in]{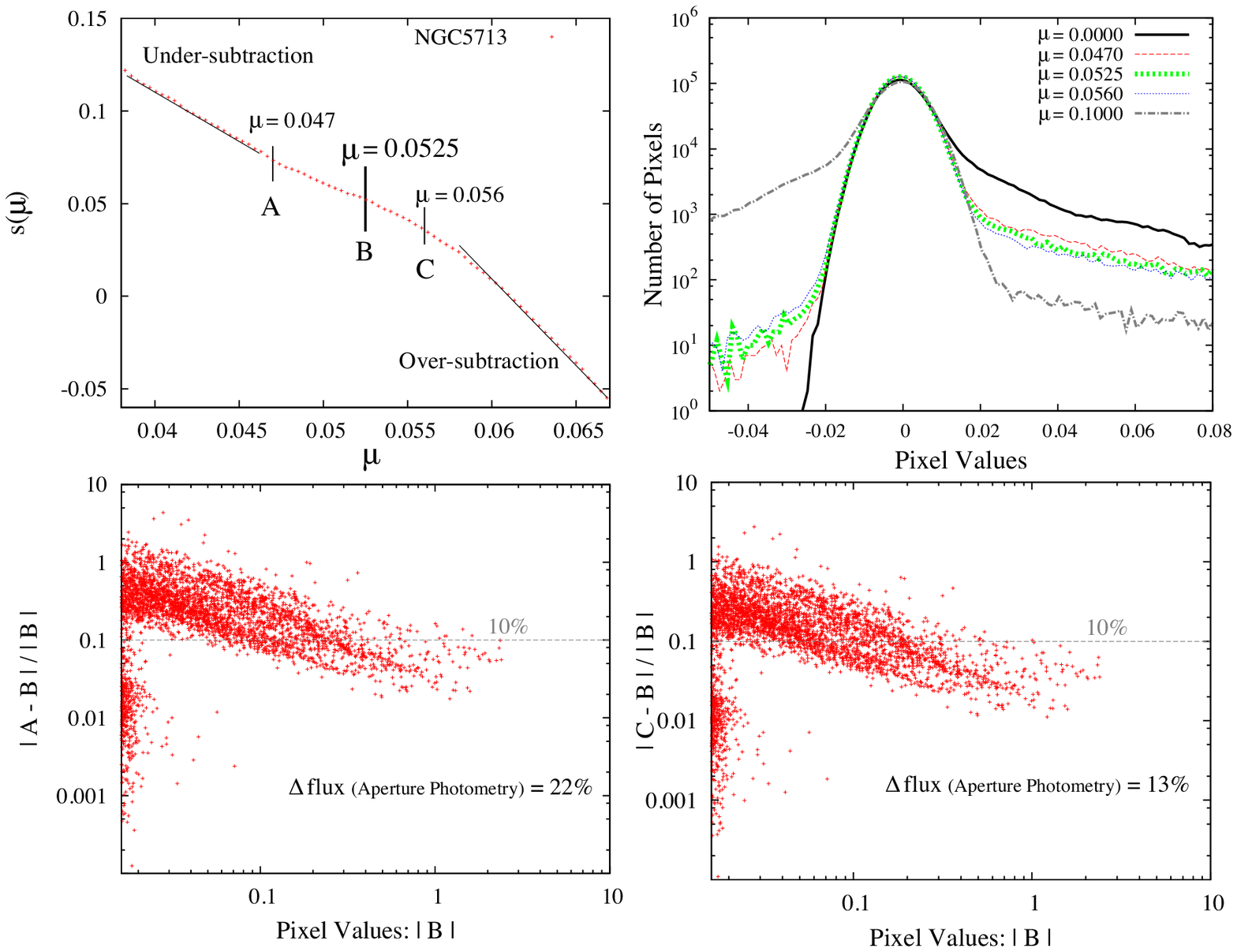}
\caption{Top-left: The zoom-in plot for the skewness function shown in Figure~\ref{fig:ngc5713first8}. 
The minimum flux ratio from stars is labelled ``A'' (and ``C'' for the minimum flux ratio). 
Top-right: The pixel histogram for each scaling factor $\mu$. The skewness in the histograms 
keep decreasing as more stellar continuum from the narrow-band image gets subtracted. 
Bottom: The relative flux difference $\frac{| A - B |}{| B |}$ between the two continuum-subtracted images on pixel-by-pixel basis 
is shown (bottom-left). The same plot for $\frac{| C - B |}{| B |}$ is also given (bottom-right). 
The $3\sigma$ level is 0.015 for the residual image B. The pixel value $| B | = 0.1$ corresponds to $20 \sigma$ level. 
The relative difference of aperture photometry for NGC5713  shown in Figure~\ref{fig:ngc5713first8} 
is  22\% (or 13\%) between A (or C) and B.
}
\label{fig:ngc5713second9}
\end{figure*}

Figure~\ref{fig:ngc5713second9} shows the zoom-in skewness function on the transition (top-left) and 
pixel histograms for various scaling factors $\mu$ (top-right). For comparison, we put  
the minimum ratio labelled ``A'' and the maximum ratio labelled ``C'' in the plot of the skewness function.   
We give the label ``B'' to our optimal choice. The transition starts near A and ends slightly further from C. 
Before and after the transition, the skewness decreases monotonically. So there are, near the transition, 
two asymptotic lines, the ``under-subtraction line" before the transition and the ``over-subtraction line" after the transition. 
If the transition is symmetric, the two asymptotic lines are parallel to each other. In this case, 
we can locate the center of the transition very accurately. The top-right panel shows the pixel histograms 
for the five residual images, the original narrow-band image ($\mu = 0$), 
A($\mu = 0.047$), B($\mu = 0.0525$), C($\mu = 0.056$), and the heavily over-subtracted 
image($\mu = 0.1$). The skewness keeps decreasing as we increase the amount of continuum-subtraction. 
Near the optimal subtraction, the decrement of skewness reduces so as to produce the transitional pattern 
in the skewness trend.

The bottom panels of Figure~\ref{fig:ngc5713second9}  show the relative flux difference $\frac{| A - B |}{| B |}$ 
(or $\frac{| C - B |}{| B |}$ ) on pixel-by-pixel basis between A (or C) and B. 
The $3\sigma$ level is 0.015 for the residual image B. So the pixel value $| B | = 0.1$ corresponds to $20 \sigma$. 
For some bright pixels, the flux differences between the two subtractions are less than 10\%. But 
for faint pixels, the fluxes are very sensitive to the choice of scaling factor $\mu$. 
For the pixels less than $20\sigma$ levels in the figure, 
the flux changes over 100\% depending on the choice of scaling factor. 
When we calculate a line ratio such as H$\beta$/\ha for diffuse gas, 
the accurate continuum subtraction becomes very important. 

The total pixel counts in the aperture for NGC5713 shown in Figure~\ref{fig:ngc5713first8} (bottom-left panel)  
are 3900 for A, 3200 for B, and 2800 for C. So the relative flux difference between A (or C) and B 
is 22\% (or 13\%). Between the two conservative subtractions A and C, the difference of aperture photometry  
range up to 30\%. So, even for aperture photometry, an inaccurate choice of continuum-subtraction can 
change the flux significantly, in our case over 10\%. 

Basically there is no mathematical proof that the center of transition is the optimal subtraction in practical applications. 
For our simulation model, the argument is mathematically correct because we set the related parameters to satisfy the perfect subtraction. 
Our empirical observations show that the ratios of stellar fluxes are located in the transition region 
and our visual inspection of the the residual image shows that a good choice for the optimal continuum--subtraction is the center of the transition region.  
We, thus, assume that the center of the transition is the optimal solution for general applications.

\subsubsection{Case 2: Space-based observations}
Basically space-based images are not much different from ground-based images. 
The only difference which matters for the skewness method is that space-based images have more resolved faint stars 
than ground-based ones, so they are more sensitive to mis-registration errors between the two images. 

\begin{figure*}[t]
\centering
\includegraphics[height=6.0 in]{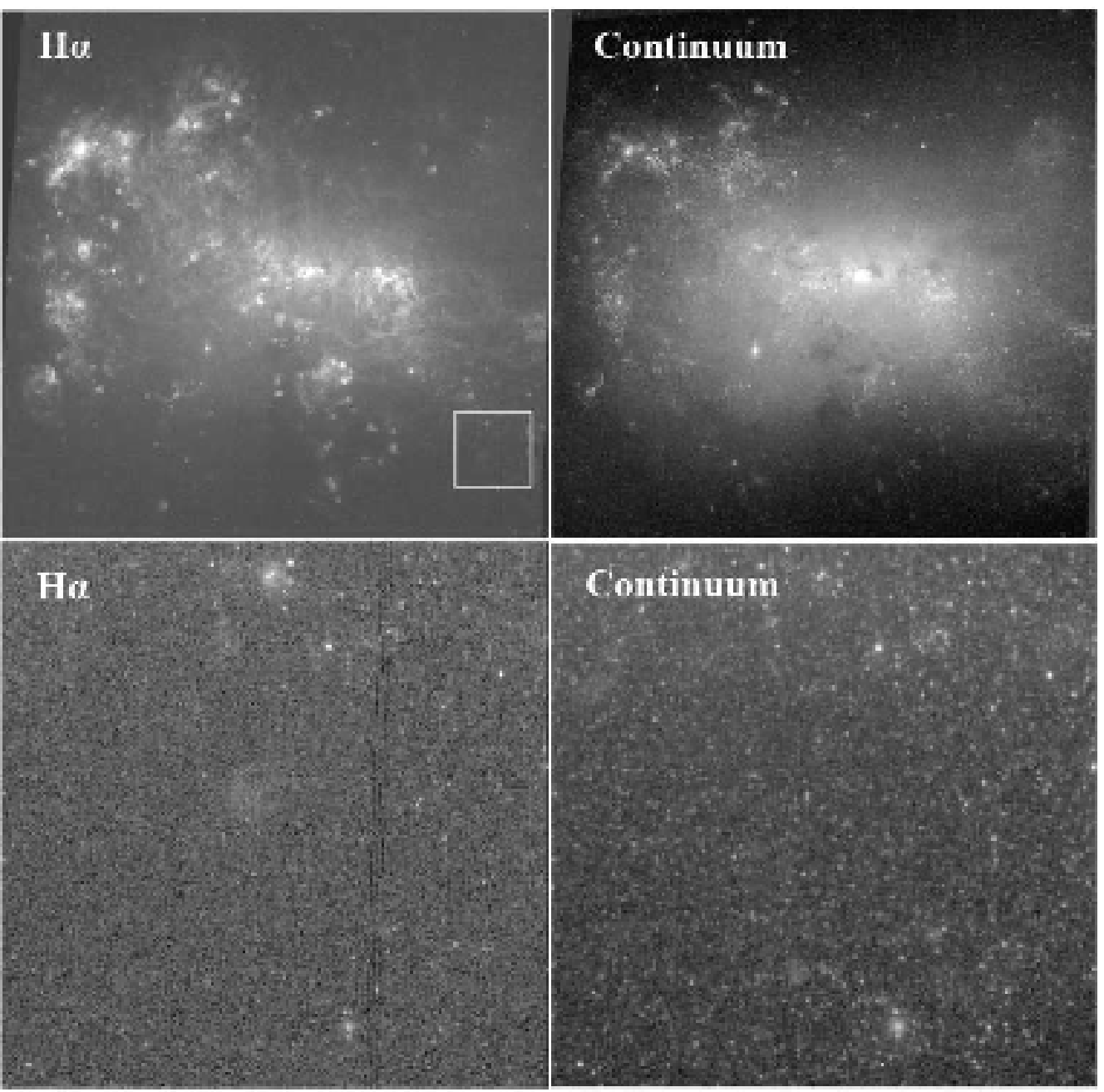}
\caption{The $H\alpha$ image (top-left) and the stellar continuum image (top-right) 
made by interpolating the two continuum images in the F814W and F555W filters of NGC4449.
Because the size of the entire image is too large and the \ha emission is bright in the starburst region, 
we choose a section of the image shown as an open square in the top-left panel. 
Bottom panels show the image section in both $H\alpha$ and continuum.  
}
\label{fig:ngc4449first10}
\end{figure*}

Figure~\ref{fig:ngc4449first10} shows the H$\alpha$ image of NGC4449 observed by the Advanced Camera for Survey(ACS) 
of the Hubble Space Telescope(HST) (top-left) and the continuum image for the H$\alpha$ image 
created by interpolating images in the two filters F814W and F555W (top-right). Since NGC4449 is a starburst galaxy, 
the nebular emission dominates over stellar objects in the H$\alpha$ image. 
Because of that, we will find a broad and asymmetric transition from the entire image and the size of the image itself is too large 
to calculate the skewness function in a relatively short amount of computing time. 
Hence, we choose the region with pixel coordinates [3400:4000, 500:1100] shown in the bottom panels in Figure \ref{fig:ngc4449first10} 
where the stellar objects dominate the total flux 
and the size is relatively small to obtain the skewness function in a short time. 

\begin{figure*}[t]
\centering
\includegraphics[height=2.7 in]{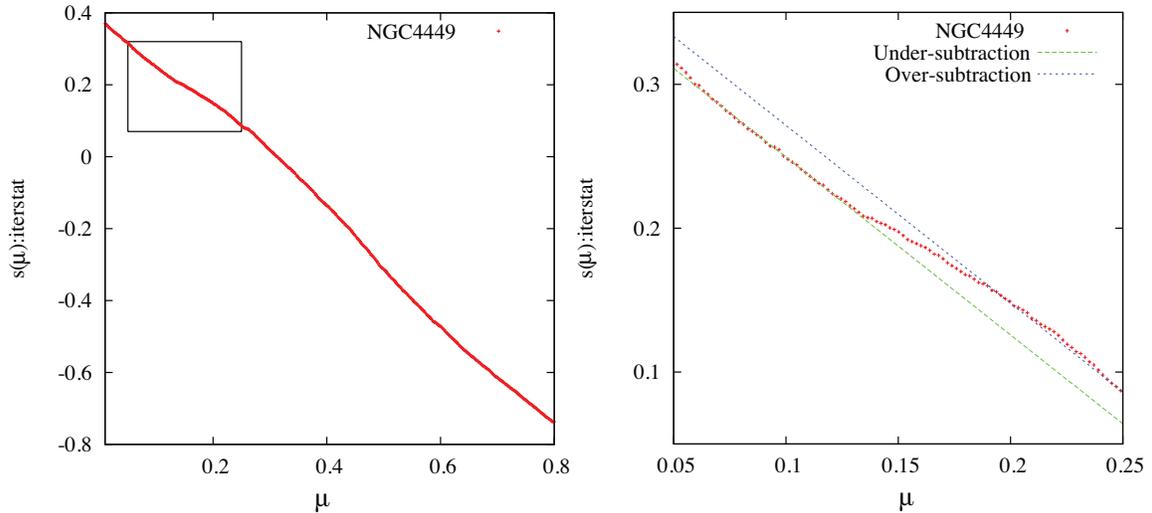}
\caption{The skewness transition for the portion image of NGC4449. 
The transition starts around $\mu=0.12$ and ends around $\mu=0.2$. 
We choose the optimal value of $\mu=0.165$.}
\label{fig:ngc4449second11}
\end{figure*}

\begin{figure*}[t]
\centering
\includegraphics[height=7.5 in]{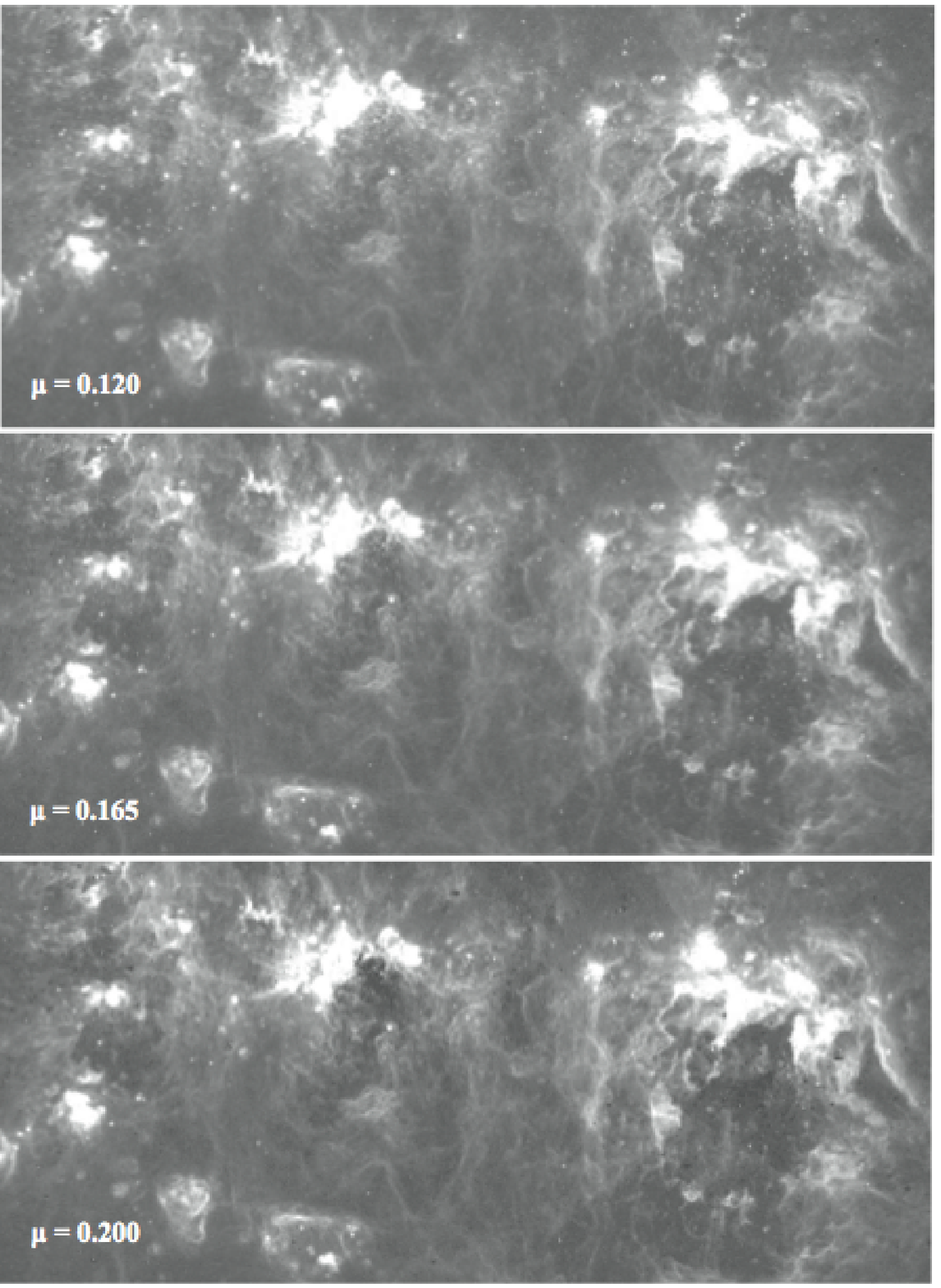}
\caption{The central region of the continuum-subtracted $H\alpha$ image with $\mu=0.12$ (under-subtraction), $\mu=0.165$ (optimal subtraction), 
and $\mu=0.2$ (over-subtraction).}
\label{fig:ngc4449third12}
\end{figure*}

The left panel of Figure~\ref{fig:ngc4449second11} shows the skewness function of the region. 
To obtain a better transition for this image we use the iterstat procedure with $5\sigma$ rejection and 10 iterations. 
The skewness function is not as smooth as the previous example of the ground-based image, because 
the small size and relatively low signal-to-noise ratio of the image 
induces noise in the skewness for each $\mu$. The right panel of Figure~\ref{fig:ngc4449second11} 
shows the transition which begins around $\mu=0.12$ and ends around $\mu=0.2$. We choose $\mu=0.165$ as 
an optimal value for the subtraction. We apply this local optimal value to the entire image. 
If the image section is large enough to represent the overall stellar population of the entire image, 
the local optimal value is a good choice for the global optimum. 
Figure~\ref{fig:ngc4449third12} shows the central region of 
the continuum subtracted images for $\mu=0.12, 0.165, 0.2$ each. 
We can observe that the skewness transition corresponds to the transition 
from under-subtraction ($\mu=0.12$) to over-subtraction ($\mu=0.2$) 
and that the optimal value chosen from the center of the transition is 
well-subtracted even at visual inspection.

Because NGC4449 is very bright in \ha, we can investigate the effect of various line emission strengths 
on the skewness transitions as shown in Figure~\ref{fig:ex3} for the simulations. 
First, for the investigation, we measure the pixel counts for the original (un-subtracted) narrow-band image. 
We consider that the pixel counts are the total flux, $\int E(x) + \sum S_i(x) dx$.   
Then we measure the pixel counts for the subtracted narrow-band image taking the optimal ratio, $\mu = 0.165$, 
found in the previous section. We consider that the pixel counts are the line emission flux, $\int E(x) dx$. 
From those, we can measure the ratio, $R \equiv \int E(x) dx/\int \sum S_i(x) dx $ (Table~1). 

\begin{figure*}[t]
\centering
\includegraphics[height=7.5 in]{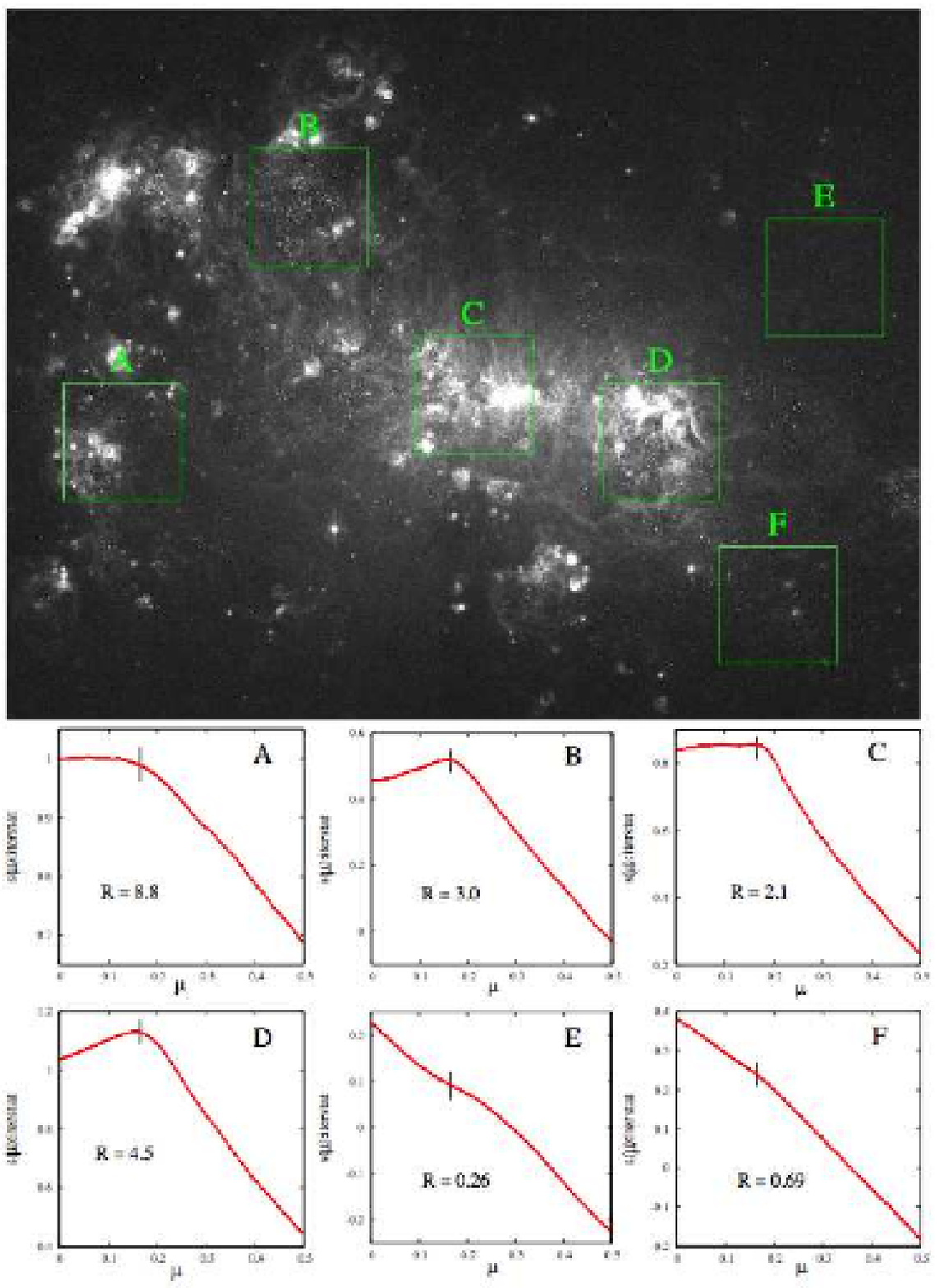}
\caption{The \ha image, the selected 6 sections named A, B, C, D, E and F (top panel) and their own skewness funtions(bottom panels). 
The vertical lines in the skewness functions indicate the location of the optimal ratio, $\mu = 0.165$. The ratio value, R, is 
the flux ratio between the total line emission and the total stellar emission as shown in Figure~\ref{fig:ex3}.  
}
\label{fig:ngc4449fourth13}
\end{figure*}

Figure~\ref{fig:ngc4449fourth13} shows the selected 6 sections in the \ha image and their skewness functions. 
The vertical lines indicate the location of the optimal ratio, $\mu = 0.165$, derived earlier from the first image region (Figure~\ref{fig:ngc4449first10}). 
We can see that our previously derived optimal ratio, $\mu = 0.165$, is consistent with 
the skewness functions of the selected sections.  
Section E, that has $R < 0.5$, shows a symmetric transition, 
for which we can locate the center easily and accurately. 
Section F has $0.5 < R < 1.0$ and shows an asymmetric trend. Sections, B, C, and D show 
the flat or slightly increasing trend before the optimal solution. After the optimal point, the skewness decreases monotonically for these regions. 
In section A, the overall trend shows a smooth decrease, although it remains similar to those of B, C, and D.  
This demonstrates that the skewness function is solely dominated by the ratio of the line--to--stellar emission. We can not find any characteristic specific to 
the stellar emission for section A.

\begin{deluxetable}{crrrr}
\tabletypesize{\footnotesize}
\tablecolumns{8}
\tablewidth{0pc}
\tablecaption{Pixel Counts for Each Section}
\tablehead{
\colhead{Region} & \colhead{Total: E+S} & \colhead{Subtracted: E} & \colhead{Remnant: S} & \colhead{E/S}
}
\startdata
A  &   6169  &   5537    &  632  &  8.8 \\ 
B  &   3106  &   2322    &  784  &  3.0 \\ 
C & 27808  & 18976  & 8832 & 2.1 \\ 
D  & 24040 & 19699 & 4341 & 4.5 \\ 
E  & 116 & 24 & 92 & 0.26 \\ 
F  & 620 & 254 & 366 & 0.69 \\ 
\enddata
\end{deluxetable}

\section{Problematic examples and applicable criteria}

We have presented the expectations and the typical applications of the skewness method in the previous sections. 
The typical transition is nearly symmetric and shows a down-slope trend. 
We choose the optimal continuum subtraction by locating the center of the transition.
However, some defects or low image quality can cause anomalous shapes of the skewness function 
so that we can not easily locate the center of transition. 
In this section, we will present cases of anomalous transitions caused by 
the presence of problematic pixels or regions in the images and show 
how those problems can be resolved. From those anomalous cases, 
we will infer criteria for the applicability of the skewness method and discuss the strengths and weaknesses of 
this method.

\subsection{Anomalous transitions}

\subsubsection{Non-uniform background}
\begin{figure*}
\centering
\includegraphics[height=8.5 in]{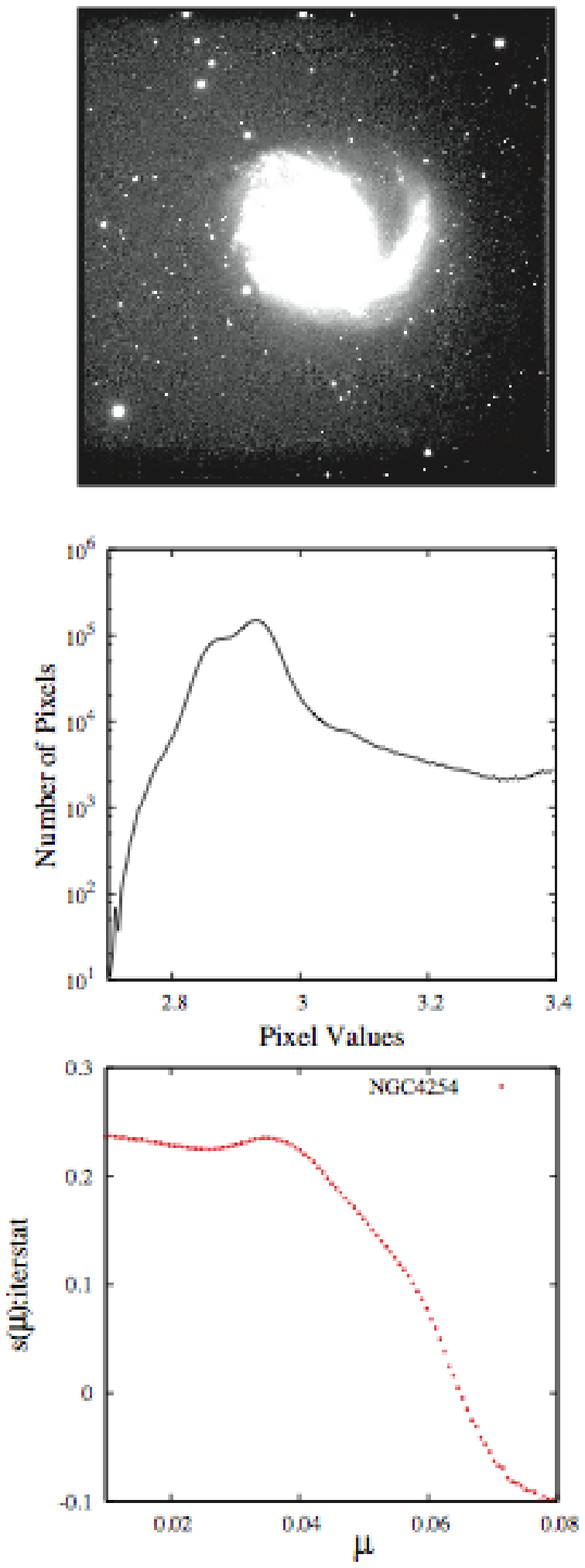}
\caption{The top panel shows the R-band image of NGC4254 and the middle panel shows the pixel-histogram 
for the image.  The noise is not Gaussian and has two peaks caused by non-uniform background. 
Because of the two peaks, the skewness function (bottom panel) shows 
the anomalous shape.
}
\label{fig:ngc4254first14}
\end{figure*}

\begin{figure*}[t]
\centering
\includegraphics[height=8.5 in]{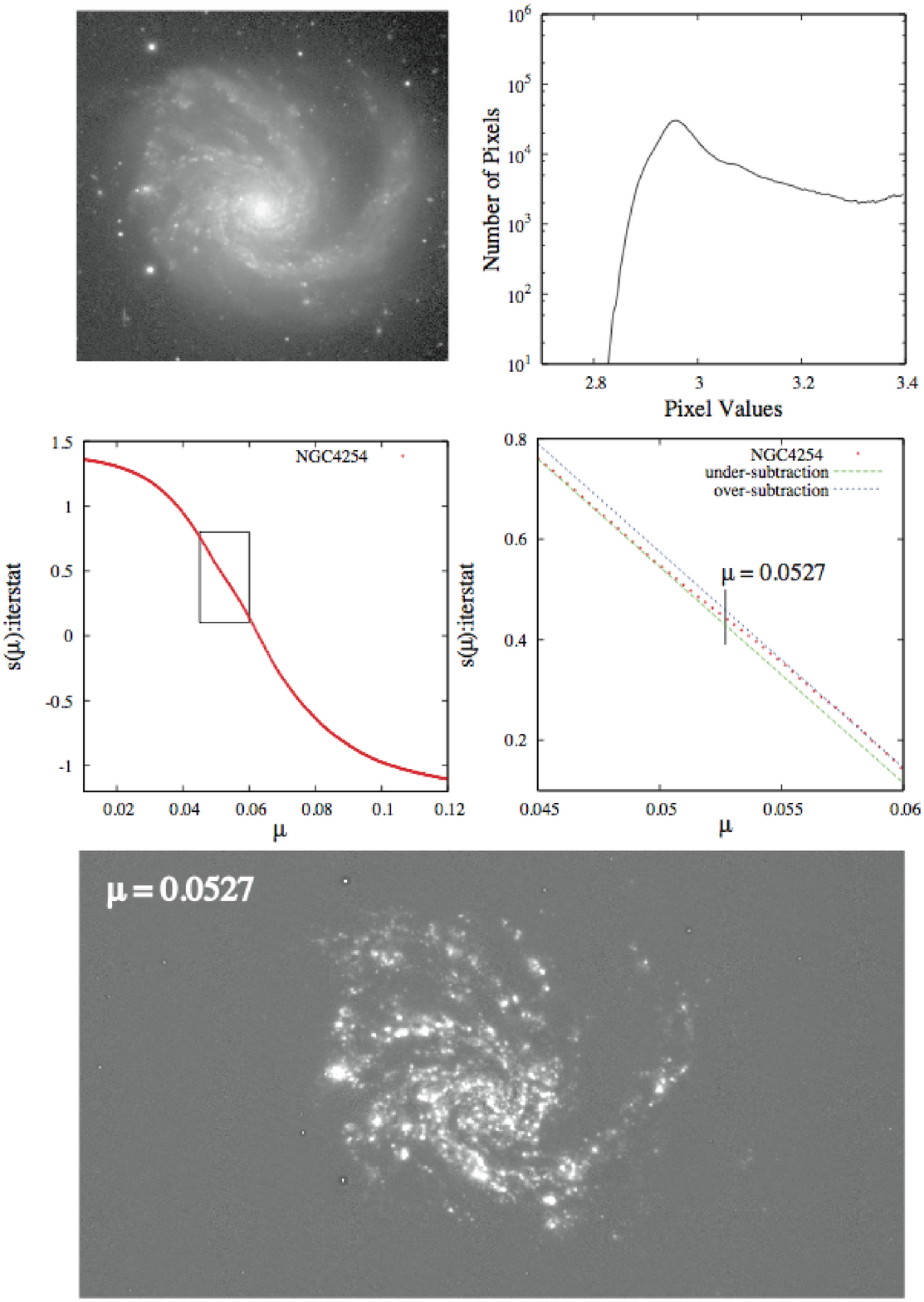}
\caption{The top-left panel shows the part of the R-band image of NGC4254 and the top-right 
panel shows the pixel-histogram for the image. The background still does not have a Gaussian noise distribution, but 
the double peak component has been removed. 
The skewness function and its zoom-in plot are shown in the middle panels. 
We choose the optimal value for subtraction as $\mu=0.0527$. The bottom panel shows the 
final subtracted image.}
\label{fig:ngc4254second15}
\end{figure*}

Figure \ref{fig:ngc4254first14} shows the R-band image of NGC4254(top) from the SINGS sample,  
its pixel histogram(middle), and its skewness function(bottom). 
The noise is not Gaussian and has two peaks caused by a non-uniform background. 
Such distorted background affects the skewness value of the image and finally produces the anomalous skewness function. 
To reduce the non-uniformity, we take the central part of the image, shown in 
Figure \ref{fig:ngc4254second15}. Even though the noise distribution is not perfectly Gaussian, the double peak component is removed. 
The middle panels in the figure show the skewness functions for this image section. The transition is now 
normal and we take the optimal value, $\mu=0.0527$. The bottom panel shows the 
continuum-subtracted H$\alpha$ image for $\mu=0.0527$ which is optimally-subtracted even at visual inspection.

\subsubsection{Indefinite sky background: the Case of Dust Lanes}

Since dust lanes of spiral galaxies appear as dark regions on a brighter background, 
some of them may be counted as sky background in pixel histograms. 
This contamination distorts the shape of sky background from its typical Gaussian distribution. 
So the off-source field is also important for our skewness transition method to secure well-defined background statistics.  
In most ground based observations the field-of-view(FOV) is large enough to cover enough off-source field. 
But the small field of view of HST observations can be problematic to obtain enough off-source field. 

\begin{figure*}[t]
\centering
\includegraphics[height=6.5 in]{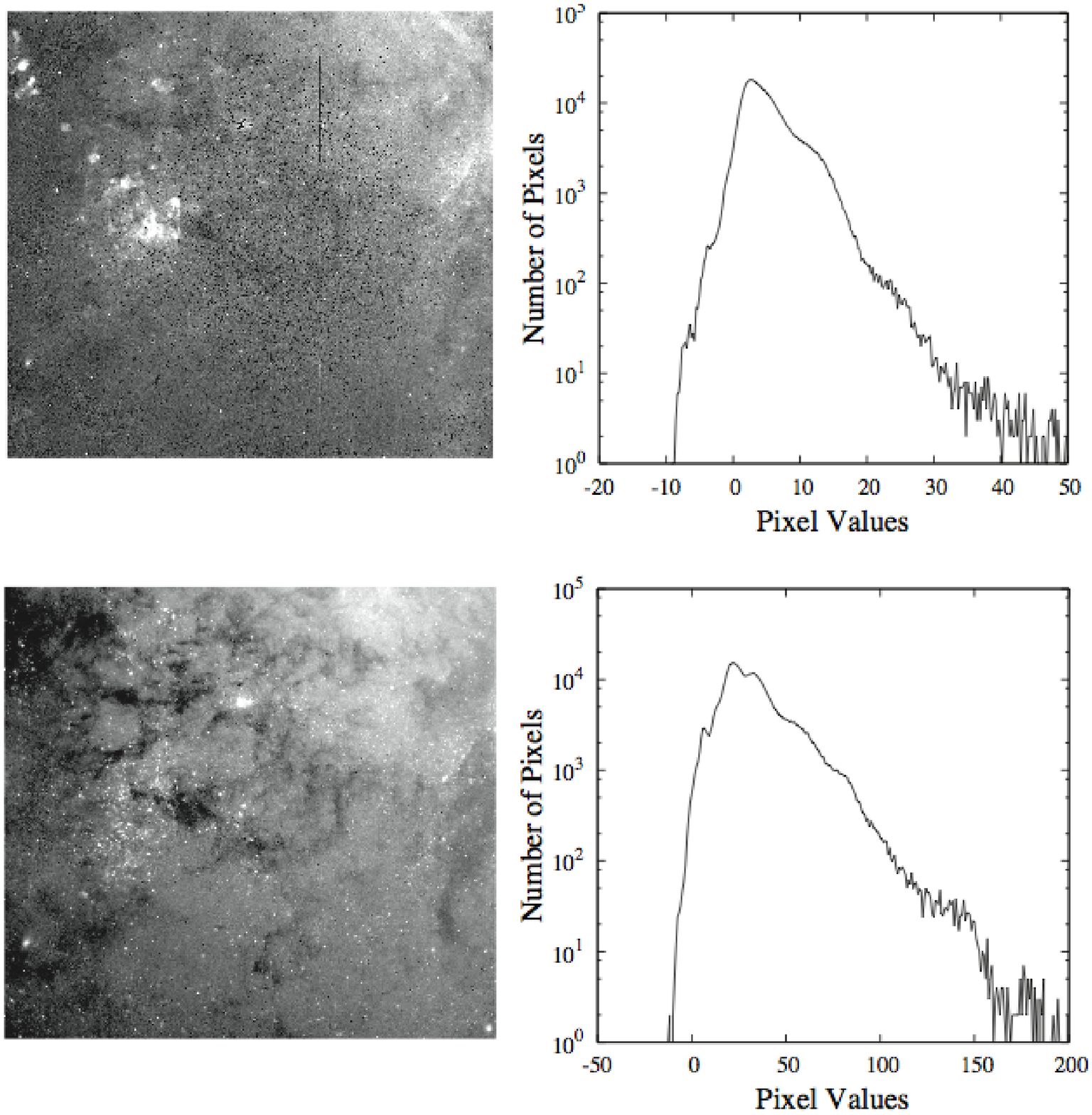}
\caption{The F658N image(top-left) and the F547M image(bottom-left) of WF2 from WFPC2 
and their pixel histograms(each right panel) for NGC4258. The pixel histograms of two images are 
irregular.}
\label{fig:ngc4258first16}
\end{figure*}

Figure \ref{fig:ngc4258first16} shows the F658N and the F547 images (left panels) of WF2 from HST/WFPC2 and 
their pixel histograms(right panels) for NGC4258. 
If the sky backgrounds are smooth and close enough to 
Gaussian distribution, we can assume that the center of the skewness transition is the optimal subtraction. 
But as irregularity increases, there is no guarantee whether the solution will be located 
at the center or not. The histograms in the figure show atypical shapes  
because of the presence of the dust lane and the lack of off-source information due to small field-of-view. 
 
\begin{figure*}[t]
\centering
\includegraphics[height=6.5 in]{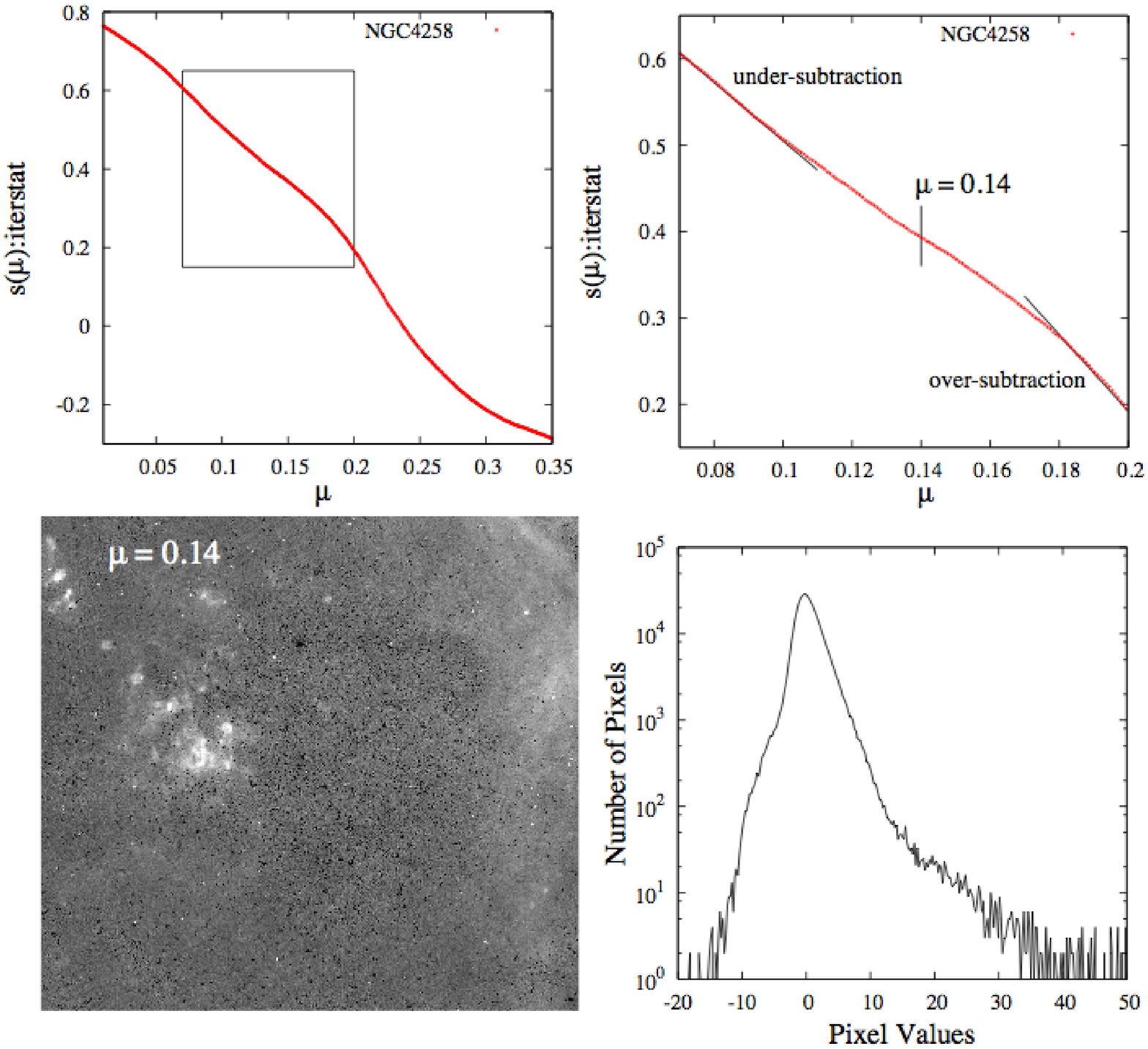}
\caption{The skewness function for NGC4258(top-left) and 
the zoom-in plot of the box from the skewness function(top-right). 
The continuum subtracted \ha image with $\mu=0.14$(bottom-left) and its pixel histogram(bottom-right).
The transition is slightly asymmetric but we can locate a possible optimal value as $\mu=0.14$. 
We can find many over-subtracted stars in the bottom-left panel, 
which implies that the transition point may not be related to the optimal subtraction.
}
\label{fig:ngc4258second17}
\end{figure*}

The top panels of Figure \ref{fig:ngc4258second17} shows the skewness function calculated from those images. 
The transition is somewhat asymmetric, but still we can locate the center of transition at 0.14. 
The problem is whether the chosen value is reliable or not. 
The bottom panels of Figure \ref{fig:ngc4258second17} show the continuum subtracted image with $\mu=0.14$ (left) 
and its pixel histogram (right). When we compare the F658N images before and after subtraction,  
Figure \ref{fig:ngc4258first16} and Figure \ref{fig:ngc4258second17} , 
we find many over-subtracted stars 
in the subtracted image; this means that the value of $\mu$ is not an optimal one.  
To avoid the over-subtraction, a better choice is to take $\mu=0.1$. 
This implies that the irregular sky backgrounds of the broad-band and narrow-band images blur the relation between 
the transition point and the optimal subtraction. For this kind of situation, the skewness method is not much more reliable 
to pick up the optimal solution than other approaches. 
But this uncertainty is an intrinsic limitation of the image itself due to the indefinite sky background 
and lack of off-source information.

\subsubsection{Saturated bright sources}
\begin{figure*}[t]
\centering
\includegraphics[height=8.5 in]{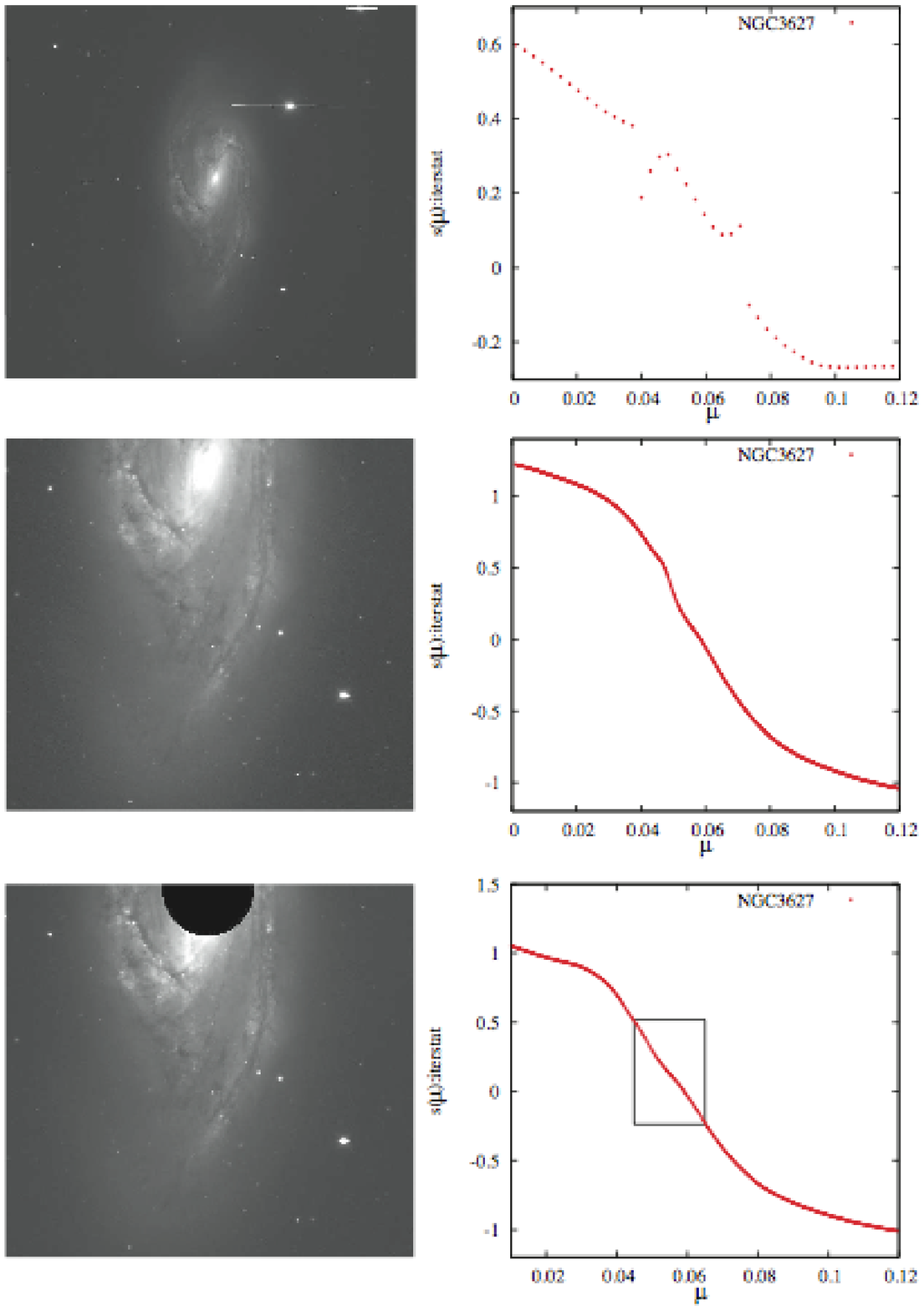}
\caption{The R-band images of NGC3627(left) and their skewness functions(right). 
The three steps (from top to bottom) increase the selectivity in the acceptable image sections, and  demonstrate the improvement in the 
skewness method when problematic regions are masked out.}
\label{fig:ngc3627first18}
\end{figure*}

Saturation of stellar objects can provide an effect on the stellar residual function and  
it will significantly affect the transition. Fig.\ref{fig:ngc3627first18} shows the skewness functions (right) 
and the R-band images (left) for NGC3627. In the top panels, the two bright stars have conspicuous horizontal spikes and 
the transition shows an unexpected up-slope trend.  
In the middle panels, we show a section of the image that avoids the two saturated stars . 
The strong up-slope trend is removed in this section, but still there is a bump around $\mu=0.05$. 
This is because of possible saturation in the galaxy bulge. We thus mask out the bulge, too.   
The final image section is shown in the bottom panels, and for this we obtain the expected transition in the skewness 
trend. 

\begin{figure*}[t]
\centering
\includegraphics[height=5.5 in]{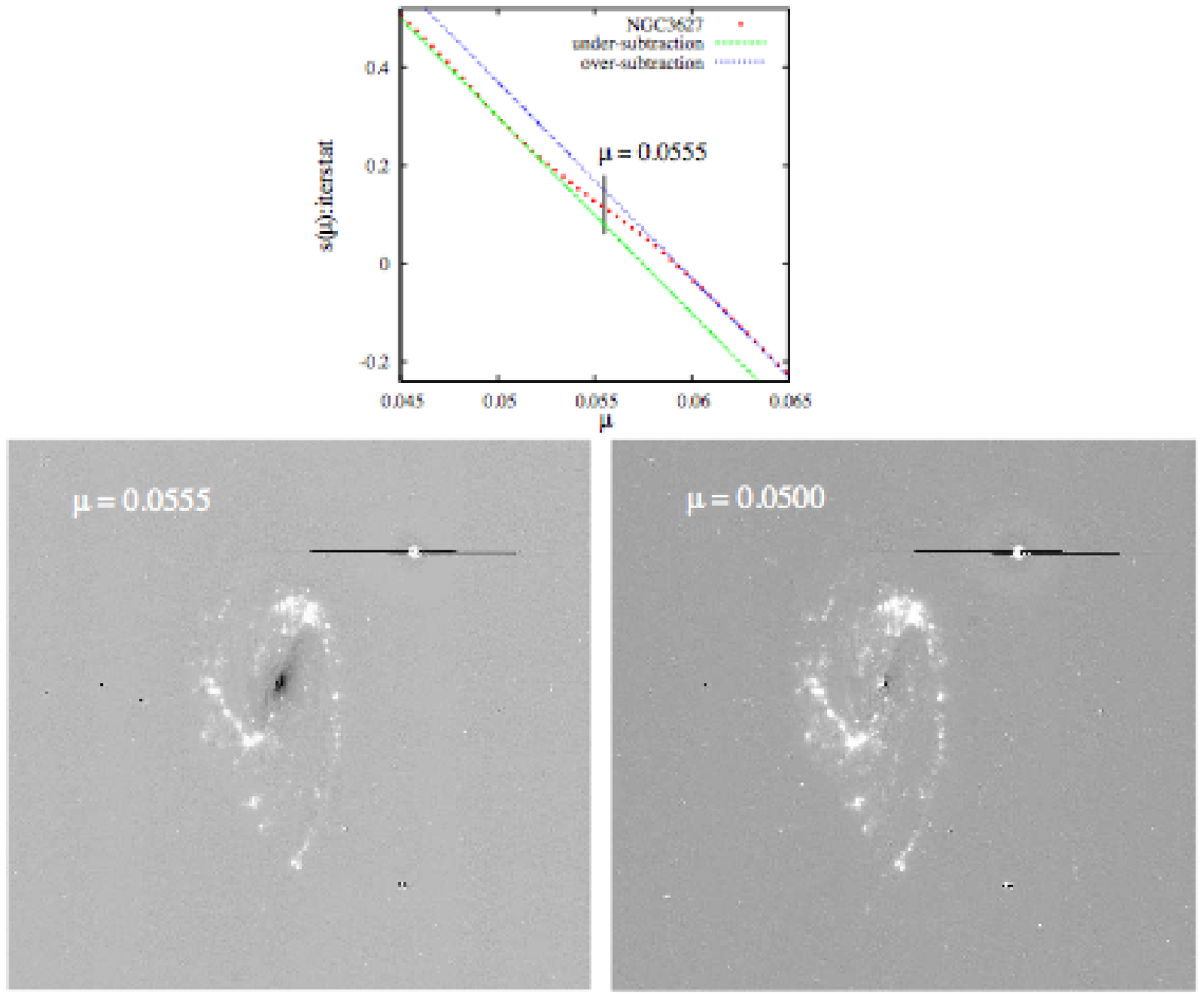}
\caption{The zoom-in plot of the skewness function in Figure~\ref{fig:ngc3627first18} (top). 
The continuum-subtracted images for $\mu=0.0555$(bottom-left) and $\mu=0.05$(bottom-right). 
We choose the optimal value $\mu=0.0555$ from the skewness function. 
}\label{fig:ngc3627second19}
\end{figure*}

The top panel of Fig.\ref{fig:ngc3627second19} shows the zoom-in plot of the transition. We take the optimal value, 
$\mu=0.0555$ from the figure. The bottom-left panel of the figure shows the 
subtracted image of $\mu=0.0555$. The bulge is over-subtracted because it was masked out in order to obtain a well--behaved skewness 
transition in the image. This shows, a posteriori, that the galaxy bulge is likely saturated in this observation.  
A visual inspection would suggest $\mu=0.05$ to avoid over-subtracting the bulge; the image resulting from this value of $\mu$ is shown  
in the bottom-right panel of the figure. However, since the skewness method is objective, and not as subjective as our 
eyes,  we choose the left panel as an optimally-subtracted image.  
These steps demonstrate the improvement on the skewness method when masking out problematic regions.

\subsection{Criteria for application}
We have presented theoretical expectations and a few applications using actual data for the skewness transition method. 
Theoretical expectations imply that the stellar emission needs to dominate in order to obtain a more symmetric 
transition and locate the optimal solution more accurately. 
And we have found that saturated bright stars or bulge, dust lane, and non-uniform sky background 
can produce anomalous behaviors in the skewness function. The effectiveness of the method can, however, be recovered  by removing/masking out 
such anomalies from the images. 

As a summary of what we have found, we can itemize the criteria of the skewness method as follows: 
\begin{enumerate}
\item Stellar emission needs to dominate over line emission ($R<1$) to obtain a more symmetric transition.
\begin{itemize} 
\item For $R < 0.5$, the transition is symmetric and we can locate the center of the transition accurately; 
\item For $0.5 < R < 1.0$, the transition becomes more asymmetric but still we can locate the transition center. 
\item For $1.0 < R$, we can still roughly locate the optimal solution using its monotonically decreasing trend after the optimal ratio. 
But when $R$ is large enough, there is no characteristic transition for the skewness function. 
\end{itemize}
\item Appropriate registrations and PSF matches are required to avoid smearing out transitions.  
\item Saturated sources need to be avoided or masked. 
\item A smooth sky background which is not too deviant from Gaussian or Poisson distributions is required.  
\end{enumerate}
Fortunately, most criteria above are not constraining requirements, in general. 
If we accept the assumption that a local optimal value from an image section is not much different from 
the global optimal value for the whole image, we can easily use image sections to satisfy the criteria for the optimal solution.

\section{Discussion}

Before we summarize our results, we discuss two important issues in this section: 
(1) self-contamination of line emission in its corresponding broad-band image; and 
(2) automation of the procedure, i.e., of finding the location of the skewness transition. 
For some cases, the target line emission also falls in the broad-band filter; 
e.g. R-band for \ha continuum subtraction. 
In general, the fraction of the line emission flux relative to the total flux in the broad--band filter is small enough that we can consider the self-contamination 
a higher order correction. 
But, in those instances in which the line emission is a non--negligible contribution, $>$10\%,  
to the broad-band image, the decontamination of the broad--band flux from the line contribution  
becomes important. A standard approach is to use an iterative subtraction method. 
The basic idea is that, if an object suffers from substantial line emission contamination in the broad-band image, 
the stellar continuum baseline is over-estimated by the amount of self--contamination. 
After subtraction, the self--contaminated object, therefore, tends to have an underestimated line flux; 
i.e. a smaller line flux estimate than a true contamination--free value. We thus keep iteratively subtracting the 
line emission and broad--band images from each other, until the fluxes stabilize to their asymptotic values, typically 
after two--three iterations. 
This iterative process can be described as : 
\begin{eqnarray}
E_0 & \leftarrow & I_N \nonumber \\
S_0 & \leftarrow & I_B \nonumber \\
E_{i+1} & \leftarrow & I_N - \mu_{i} S_{i} \nonumber \\
S_{i+1} & \leftarrow & I_B - E_{i+1}, \nonumber 
\end{eqnarray}
where $I_N$ is the original narrow-band image, $I_B$ the original broad-band image, $E_{i}$ the i-th line emission image, 
$S_{i}$ the i-th stellar continuum image, and $\mu_{i}$ the i-th optimal ratio of continuum subtraction. 
The two original images, $I_N$ and $I_B$, need to be scaled to the same units to ensure that the subtraction, $I_B - E_{i+1}$, is self--consistent. 
During the iterations, if we can take the best ideal $\mu_i$ for each step, 
the two sequences of $\{E_i\}$ and $\{S_i\}$ should converge to the contamination-free values. 
The consistent choices of $\mu_i$s, hence, are important to guarantee the convergence. 
We applied this iteration method to decontaminate \oiii $~$and \hb $~$from F555W in Hong et al. (2011). 
We compared the iterated F555W with the F547M image which is the filter free from line contamination. 
From this study, we have shown that the iteration method works and  
our skewness transition method is good enough to provide accurate subtractions to obtain convergent values. 
However, it is still possible for some images with poor quality to fail to converge during iterative subtractions. 

The typical symmetric transition (or near-symmetric transition; R $< 0.5$) has an inflection point 
where the optimal solution is (See Figure~\ref{fig:ex3}). This is a fundamental feature 
to build an automatic method for continuum subtraction. 
But the problem is, as described in \S 3, that we have observational anomalies which are generally handled manually. 
Therefore, the automation of the continuum subtraction procedure suffers from practical issues. 
We can implement a code to find the inflection point and infer the corresponding R value to check whether the 
inflection point is derived from a symmetric transition or not. 
Auxiliary codes which can recognize anomalies in the transition and select anomaly--free image sections are, however, needed to avoid 
problematic regions.  
There can be many solutions to deal with this practical issue such as genetic algorithm and pattern recognition. 
One option is to choose random sections in an image and calculate the transition for each section. The results from the image sections could then 
be discarded or merged  according to the quality of each solution. The need for automatic pipelines to handle huge amounts of data in 
astronomy will ultimately drive the requirements for the automation of this method, via optimized routines and/or interfaces, in order to increase its 
practical use. 

\section{Summary}

We have presented a quantitative method to determine the optimal stellar continuum subtraction from a narrow-band image. 
The skewness of the residual image shows a transitional feature and we have found that, from our simulations, 
the transition is related to the optimal value of the subtraction. If the stellar emission dominates over the line emission in the narrow-band image, 
the transition is symmetric and the optimal solution is located at the center of the transition; hence a very accurate subtraction can be 
achieved. 

Uncertainties brought by nonuniform sky background and saturation of bright stars or bulges can produce 
anomalous transitions. If we can identify ``well-behaved'' (i.e., anomaly--free) image sections that satisfy the applicable criteria (section 3.2), 
we recover the conditions for optimal subtraction.  
This method is objective and quantitative, and it is statistically complete since it use all the pixels in the image. 
We can also obtain a better precision than visual subtraction 
if the images have sharp and symmetric transitions. This accuracy is important especially for diffuse gas emission, for which an inaccurate stellar 
continuum subtraction can produce over 100\% variation in the measured flux.  By choosing well--behaved sections of images and assuming that the 
local optimal value is not much different 
from the global one, we can achieve high accuracy ($<$10\% error on the global flux) for most astronomical images. 
\clearpage


\begin{thebibliography}{}
\bibitem{calzetti} Calzetti, D., Kennicutt, R. C., Engelbracht, C. W., et al. 2007, ApJ, 666, 870
\bibitem{helou} Helou, G., Roussel, H., Appleton, P., et al. 2004, ApJS, 154, 253
\bibitem{hong} Hong, S., Calzetti, D., Dopita, M. A. et al. 2011, ApJ, 731, 45
\bibitem{kennicutt2003} Kennicutt, R.C., Armus, L., Bendo, G., Calzetti, D., Dale, D.A., Draine, B.T., Engelbracht, C.W., et al. 2003, PASP, 115, 928
\bibitem{kennicutt} Kennicutt, R. C., Lee, J. C., Funes, S. J., Sakai, S., \& Akiyama, S. 2008, ApJS, 178, 247
\end{thebibliography}
\end{document}